\documentclass[conference]{IEEEtran}
\IEEEoverridecommandlockouts
\usepackage{cite}
\usepackage{amsmath,amssymb,amsfonts}
\usepackage{algorithmic}
\usepackage{graphicx}
\usepackage{textcomp}
\usepackage{xcolor}
\usepackage[hyphens]{url}

\usepackage{mathtools}
\usepackage{multirow}
\usepackage{xspace}
\usepackage{soul}
\usepackage{tikz}
\usepackage{gensymb}
\usepackage{booktabs}
\usepackage{tikz}
\usepackage[subtle]{savetrees}

\def\BibTeX{{\rm B\kern-.05em{\sc i\kern-.025em b}\kern-.08em
    T\kern-.1667em\lower.7ex\hbox{E}\kern-.125emX}}

\newcommand{\ACC}{\textit{Mirage}\xspace}

\newcommand{\microarchitecture}{micro-architecture\xspace}
\newcommand*\circled[1]{\tikz[baseline=(char.base)]{
            \node[shape=circle,draw,inner sep=1.1pt] (char) {#1};}}
\title{\ACC: An RNS-Based Photonic Accelerator for\\
DNN Training 
\author{\IEEEauthorblockN{Cansu Demirkiran}
\IEEEauthorblockA{
\textit{Boston University}\\
Boston, MA, USA \\
cansu@bu.edu
}
\and
\IEEEauthorblockN{Guowei Yang}
\IEEEauthorblockA{
\textit{Boston University}\\
Boston, MA, USA \\
guoweiy@bu.edu 
}
\and
\IEEEauthorblockN{Darius Bunandar}
\IEEEauthorblockA{
\textit{Lightmatter}\\
Boston, MA, USA \\
darius@lightmatter.co
}
\and
\IEEEauthorblockN{Ajay Joshi}
\IEEEauthorblockA{
\textit{Boston University}\\
Boston, MA, USA \\
joshi@bu.edu
}

}
}
\begin{document}
\newpage
% \input{reviews_2}
% \maketitle
\microtypesetup{tracking=false}
\maketitle
\microtypesetup{tracking=true}
\setcounter{page}{1}
\thispagestyle{plain}
\pagestyle{plain}

%%%%%% -- PAPER CONTENT STARTS-- %%%%%%%%

\begin{abstract}
Photonic computing is a compelling avenue for performing highly efficient matrix multiplication, a crucial operation in Deep Neural Networks (DNNs). 
While this method has shown great success in DNN inference, meeting the high precision demands of DNN training proves challenging due to the precision limitations imposed by costly data converters and the analog noise inherent in photonic hardware.
This paper proposes \ACC, a photonic DNN training accelerator that overcomes the precision challenges in photonic hardware using the Residue Number System (RNS). 
RNS is a numeral system based on modular arithmetic---allowing us to perform high-precision operations via multiple low-precision modular operations. 
In this work, we present a novel \microarchitecture and dataflow for an RNS-based photonic tensor core performing modular arithmetic in the analog domain. 
By combining RNS and photonics, \ACC provides high energy efficiency without compromising precision and can successfully train state-of-the-art DNNs achieving accuracy comparable to FP32 training. 
Our study shows that on average across several DNNs when compared to systolic arrays, \ACC achieves more than $23.8\times$ faster training and $32.1\times$ lower EDP in an iso-energy scenario and consumes $42.8\times$ lower power with comparable or better EDP in an iso-area scenario. 
% \vspace{-0.02in}
\end{abstract}

\section{Introduction}
\label{sec:intro}

Photonic computing has shown remarkable promise in accelerating Deep Neural Network (DNN) inference by enabling high throughput and energy-efficient matrix/vector operations~\cite{shen2017deep, xu202111, shiflett2021albireo,demirkiran2023electro, mehrabian2018pcnna, sunny2021crosslight, liu2019holylight, peng2020dnnara, shiflett2020pixel}. 
However, photonic computing---similar to other analog computing methods---suffers from precision issues, making it unsuitable for DNN training.

Photonic DNN accelerators are typically centered around a matrix-vector multiplication (MVM) unit based on Mach Zehnder Interferometers (MZIs)~\cite{shen2017deep,demirkiran2023electro}, Micro-Ring Resonators (MRRs)~\cite{mehrabian2018pcnna, sunny2021crosslight, liu2019holylight, peng2020dnnara}, or a combination of the two~\cite{shiflett2021albireo, shiflett2020pixel}. 
These units perform a set of dot products in parallel in the analog domain for executing general matrix-matrix multiplication (GEMM) operations. 
Operations in the MVM unit require switching between digital and analog domains before and after each analog operation.
%as data can only be stored digitally.
These conversions are performed by using digital-to-analog and analog-to-digital converters (DACs and ADCs).
In the analog domain, the values can only be represented as fixed-point (FXP) numbers whose precision is limited by the precision of DACs and ADCs, as the values are encoded in physical properties. 

To better understand this limitation, assume an input vector and a weight vector---that are to be multiplied in the analog domain---pass through DACs with $b_i$-bit and $b_w$-bit precision, respectively. 
The DACs convert these values into $b_i$-bit and $b_w$-bit signed integers. 
A dot product between two $h$-long vectors with $b_w$-bit and $b_i$-bit elements then produces a result with $b_\text{out} {=} b_i {+} b_w {+} \log_2(h){-}1$ bits of information.
To capture this output fully, an ADC with bit precision $b_{\text{ADC}} \geq b_\text{out}$ is needed.
%When this output passes through an ADC with bit precision $b_{\text{ADC}}$, only $b_{\text{ADC}}$ bits out of $b_\text{out}$ bits can be captured unless $b_{\text{ADC}} \geq b_\text{out}$.  
Unfortunately, the energy consumption of ADCs increases \emph{exponentially} with bit precision and can easily make the high energy efficiency promise of photonic computing for DNNs no longer feasible. 
To avoid this, the common practice is to use ADCs with $b_{\text{ADC}} {<} b_\text{out}$ and lose $b_\text{out} {-} b_{\text{ADC}}$ bits of information on every dot product result~\cite{rekhi2019analog}. 
In addition, vector size $h$, which is determined by the size of the photonic core, is typically smaller than the matrix sizes in large DNN layers. %a full GEMM operation cannot be executed at once. 
This requires the matrices to be tiled into multiple smaller blocks and the GEMM operation to be executed tile-by-tile. 
Each tiled-MVM operation produces a partial output to the final GEMM output where the aforementioned information loss is induced on every partial output---causing the errors to accumulate while composing the final GEMM output.

While some simple DNNs are more resilient to noise, typically, as the DNN gets larger and the task becomes more complex, a degradation in accuracy is observed~\cite{rekhi2019analog, demirkiran2023blueprint}.
This degradation can be quite drastic even for DNN inference, especially for higher $h$.
Furthermore, DNN training is often more sensitive to quantization noise than inference, as gradient calculation requires a relatively higher dynamic range.
Therefore, the scope of DNN acceleration with photonic hardware has been limited to DNN inference except for a few works~\cite{bandyopadhyay2023photonic, Filipovich:22, pai2023experimentally, hughes2018training, zhang2021efficient} that focused on training very small DNN models and simple tasks. 
Generally speaking, training state-of-the-art DNN models using photonic compute cores has been a far-fetched goal due to the limited precision of analog operations. 

In this paper, we present \ACC, a precise photonic accelerator for DNN training. 
\ACC leverages the Residue Number System (RNS), a numeral system based on modular arithmetic, to perform high-precision analog operations in the photonic hardware.
In RNS, numbers are represented as a set of integer residues for a set of selected co-prime moduli---reducing the bit-width of the operands.
%As RNS is closed under addition and multiplication, a full GEMM operation can be performed in the RNS space by using the low-bit-width residues. 
Essentially, RNS simplifies high-precision operations into multiple low-precision operations and enables high-precision arithmetic despite using low-precision data converters. 
As a result, we can take advantage of the high-speed and energy-efficiency opportunities of photonic computing while achieving high DNN accuracy.

GEMM operations in the RNS space require modular dot products. 
\ACC employs a novel micro-architecture for performing modular multiply-accumulate (MAC) operations in the photonic core by leveraging the fact that the optical phase loops around at every $2\pi$---which is effectively a modulo $2\pi$ operation. 
We use phase shifters to perform modular multiplications between two operands that are encoded in the applied voltage and the length of the phase shifters.
%This enables us to apply an amount of phase shift that is proportional to the multiplication of two operands on the optical signal. 
%By utilizing separate phase shifters for each digit of the operand and choosing to go through or bypass the phase shifter in each digit, we can manipulate the total length of the phase shifter in each multiplication. 
We cascade multiple modular multipliers to accumulate the multiplication results encoded in the optical phase to perform modular dot products. 
A set of modular dot product units (MDPUs) construct a modular MVM unit (MMVMU).
We deploy a separate MMVMU for each $n$ moduli to perform $n$ modular MVMs in parallel. 
%We perform Binary Number System (BNS)-to-RNS and RNS-to-BNS conversions before and after each MVM, respectively, and perform the rest of the operations (e.g., nonlinearities, embeddings, etc.) digitally in FP32. 
We leverage the block floating point (BFP) format to obtain a high dynamic range while enabling integer operations in the photonic core using the mantissa bits of BFP values that share an exponent.
%a group of elements sharing an exponent. 
%To minimize the hardware cost of the BNS-RNS and RNS-BNS conversions, we use a special moduli set that allows fast and efficient conversions~\cite{hiasat2019residue}.

%To evaluate our design, we first perform a sensitivity analysis to choose the optimal BFP configuration, size and number of photonic MVM arrays, and the dataflow in the photonic core that achieves maximum energy efficiency while maintaining high DNN accuracy. 
To evaluate our design, we first perform a sensitivity analysis to choose the optimal BFP configuration and make micro-architecture decisions that lead to maximum performance in \ACC.
We then compare the training performance of \ACC against systolic arrays that support FP32 and several other more efficient data formats.
%While not optimized for inference, we show the applicability of \ACC to DNN inference and compare its performance against existing photonic and electronic DNN inference accelerators. 
%such as INT8 and INT12, bfloat16~\cite{bfloat16}, HFP8~\cite{sun2019hybrid}, and FAST~\cite{zhang2022fast}.

Our contributions in this work are as follows: 
\begin{itemize}
    
    \item We propose an RNS-based computing model and dataflow for DNN training that combines BFP and RNS to achieve high accuracy in analog photonic hardware. 
    \item We introduce a new photonic MAC unit design that efficiently performs modular operations using the optical phase, which enables RNS-based arithmetic using photonic hardware.
    \item Using the photonic MAC units as a building block, we architect a novel RNS-based photonic DNN training accelerator, {\ACC}. To our knowledge, {\ACC} is the first-ever photonic training accelerator that can successfully train real-world practical DNN models. 
\end{itemize}

Our evaluation shows that {\ACC} can train state-of-the-art DNNs in various tasks
%in image classification, object detection, and translation tasks 
and achieve validation accuracy comparable to FP32 training.
We show that on average across several DNNs, \ACC achieves more than $23.8\times$ faster training and $32.1\times$ lower Energy-Delay Product (EDP) in an iso-energy scenario and consumes $42.8\times$ lower power with comparable EDP in an iso-area scenario when compared to systolic arrays using the best-performing data format. 

% \ACC achieves more than $..\times$ better inferences per second per Watt (IPS/W) than existing photonic DNN inference accelerators and more than $... \times$ better IPS/W than electronic accelerators.
% We show that when we conduct an iso-area hardware cost analysis, on average among different DNNs, {\ACC} completes the same training task {\color{red}\textit{xx.yy}} $\times$ faster by consuming {\color{red}\textit{xx.yy} $\times$} less energy compared to a systolic array with FP32 MAC units. 
% When more efficient data formats are used, {\ACC} is at least {\color{red}\textit{xx.yy} $\times$} faster and consumes {\color{red}\textit{xx.yy} $\times$} less energy. 
%\cansu{Add LLM results.} 
%GPU while consuming \textit{xx.yy} $\times$ less power. 

\section{Background}
\label{sec:background}
\vspace{-0.1in}
\subsection{DNN Training}

% A DNN consists of a sequence of $L$ layers.
DNN training consists of two main steps: forward pass and backward pass. 
In a DNN, the input $X$ to the $({\ell}{+}1)$-th layer during a forward pass is the output generated by the previous ($(\ell)$-th) layer:
% \vspace{-0.05in}
\begin{equation}
    X^{(\ell+1)} = f^{(\ell)} \big( W^{(\ell)} X^{(\ell)} \big),
    \label{eq:nn}
\end{equation}
where $ O^{(\ell)} = W^{(\ell)} X^{(\ell)}$ is a GEMM operation, $W^{(\ell)}$ is the weight matrix and $f^{(\ell)}(\cdot)$ is the nonlinear function of the $(\ell)$-th layer.
After the forward pass, a loss value $\mathcal{L}$ is calculated using the output collected in the forward pass and the ground truth. 
The gradients of the activations and DNN parameters with respect to $\mathcal{L}$ for each layer are calculated by performing a backward pass after each forward pass:
% \vspace{-0.05in}
\begin{equation}
    \frac{\partial{\mathcal{L}}}{\partial {X^{(\ell)}}} =  {{W^{(\ell)}}^T} \frac{\partial{\mathcal{L}}}{\partial {O^{(\ell)}}},
    \label{eq:back-x}
\end{equation}
\begin{equation}
    \frac{\partial{\mathcal{L}}}{\partial {W^{(\ell)}}} =  \frac{\partial{\mathcal{L}}}{\partial {O^{(\ell)}}} {X^{(\ell)}}^T.
    \label{eq:back-w}
\end{equation}
Using these gradients ${\frac{\partial{\mathcal{L}}}{\partial {W^{(\ell)}}}}$, i.e., $\Delta W^{(\ell)}$, the DNN parameters are updated in each iteration $i$ as:
% \vspace{-0.05in}
\begin{equation}
W^{(\ell)}_{i+1} = W^{(\ell)}_i - \eta \Delta W^{(\ell)}_i,
    \label{eq:w-update}
\end{equation}
using a step size $\eta$ for the stochastic gradient descent (SGD) optimization algorithm. 
Essentially, for each layer, one GEMM operation is performed during the forward pass and two GEMM operations are performed during the backward pass. 

\subsection{Data Formats for DNNs}
\label{sec:backgr-number-formats}
The efforts to optimize DNN training or inference mainly revolve around improving the efficiency of MAC operations and matrix multiplications.
To accelerate MAC operations, scalar quantization and FXP multiplications have been extensively explored for DNN inference and training~\cite{courbariaux2014training, banner2018scalable, hubara2017quantized, zhou2016dorefa, gupta2015deep, jacob2018quantization}.
In addition, special FP formats have been proposed to improve the hardware performance over FP32 and FP16 formats. 
Examples include Brain Float (BFLOAT16)~\cite{bfloat16}, HFP8~\cite{sun2019hybrid}, and TensorFloat~\cite{tensorfloat}.

In our work, we use the BFP format, which provides a middle ground between FXP and FP formats.
BFP has been used for DNN inference~\cite{song2018computation, basumallik2022adaptive, darvish2020pushing} and training~\cite{drumond2018training, zhang2022fast} as it is less costly than the FP formats and achieves better accuracy than the FXP formats with the same bit-width.
BFP format splits tensors into groups and assigns an exponent to each group that is shared by the elements within the group. 
This representation allows integer operations between groups using only sign and mantissa bits while preserving the dynamic range through a shared exponent.

In the analog domain, the lack of FP arithmetic and the lack of support for large-bit-width FXP numbers limit the range of applications that can benefit from analog computing.
Combining BFP with analog hardware is a promising solution as BFP enables low-precision integer operations while providing a wider dynamic range than conventional integer arithmetic.
This idea of using BFP formats for analog computing was first proposed by Basumallik et al.~\cite{basumallik2022adaptive} for DNN inference. 
In our work, we combine BFP and RNS to perform DNN training in photonic hardware. 

\subsection{Bit Precision in Conventional Analog Cores}
\label{sec:backgr-bit-prec}
\begin{figure}
    \centering
    \includegraphics[width=\linewidth]{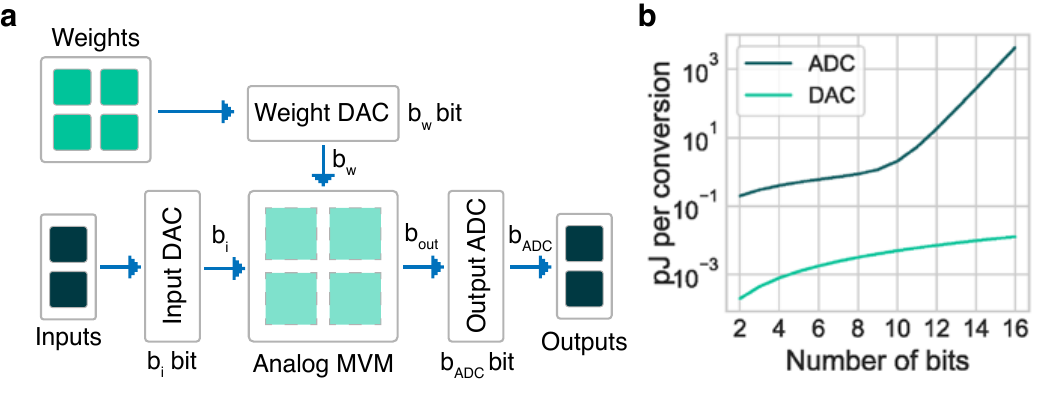}
    % \vspace{-0.15in}
    \caption{(a) Dataflow for a conventional analog core. (b) Energy consumption per conversion in ADCs and DACs with varying bit precision. The energy per conversion numbers are estimated using equations formulated by Murmann~\cite{murmann21mixed}.
    }
    \label{fig:conv_en_acc}
    % \vspace{-0.25in}
\end{figure}

Fig.~\ref{fig:conv_en_acc}(a) shows a diagram for the dataflow in a conventional analog core performing a single MVM operation. 
Independent of the technology, in an analog MVM core, inputs and weights in a DNN layer are passed through DACs and are encoded in an analog property (e.g., phase, amplitude, etc.).  
After analog dot products are performed, the output data are passed through ADCs. 
Here, the precision of the analog operation is determined by (1) the precision of DACs, (2) the precision of ADCs, and (3) the signal-to-noise ratio (SNR) during the analog operations. 

A dot product between $b_\text{in}$-bit input and $b_w$-bit weight---both $h$-long vectors and encoded by DACs---results in $b_\text{out} {=} b_\text{in} {+} b_w {+} \text{log}_2(h){-}1$ bits of information. 
For example, for 8-bit DACs, the output will require more than 16 bits, calling for an ADC with $b_\text{ADC}{\geq} 16$ to ensure no information loss. 
Fig.~\ref{fig:conv_en_acc}(b) shows the approximate energy consumption per conversion for ADCs and DACs with different bit precision~\cite{murmann21mixed}.  
As seen in the figure, ADC energy consumption is much higher (two orders of magnitude) than DAC energy consumption. 
In addition, ADC energy consumption increases exponentially with bit precision---roughly $4\times$ higher energy per conversion for each additional bit. 
For the abovementioned 8-bit example, a single A-to-D conversion would require ${\geq}1$ nJ energy. 
Considering the low energy consumption of the MAC operations performed in the analog domain (tens-to-hundreds of fJ/MAC), high-precision ADCs can easily dominate the total energy consumption.
In addition to data converter precision, the SNR required to ensure the integrity of the analog operations also increases exponentially with the increasing bit precision, in turn calling for higher power consumption and posing a limitation to achievable precision in analog cores.  
% In addition to data converter precision, the SNR 
% requirements in an analog core also pose challenges.
% As bit precision increases, a higher SNR in the analog core should be maintained to ensure the integrity of the analog operations. 
% SNR requirement increases exponentially with increasing bit precision, in turn calling for higher power consumption and posing a limitation to achievable precision in the analog core. 

% {\color{magenta}
While such high-precision analog operations are impractical, using low precision causes information loss on partial outputs.
Prior works~\cite{rekhi2019analog, demirkiran2023blueprint} show that this information loss due to low-precision ADCs can cause a drastic accuracy loss for inference, even for 8-bit analog operations---which is a widely accepted bit precision for DNN inference in digital hardware.
This difference between analog and digital hardware stems from that in digital hardware, the quantization step is typically done after scaling by a static value (e.g., a pre-determined constant) or a dynamic value dependent on the input (e.g., the maximum value in a tensor or a shared exponent as in BFP) that helps minimize the quantization effects and preserve the dynamic range.
In analog hardware, while this scaling can be done at the quantization step before processing a layer (before DACs), the output of the analog MAC operations is quantized by the ADCs without any scaling in the analog domain---causing a more significant loss than the pre-layer quantization.  
Here, a higher $h$ increases the required bit precision to capture the partial outputs ($b_\text{out}$)---resulting in higher information loss on the partial outputs and in turn, more severe accuracy degradation, especially for large DNNs handling complex tasks. 
%The accuracy loss gets more profound as the vector size $h$ increases, especially when the DNN is larger and handles more complex tasks.
% However, reducing $h$ restricts the parallelism within the photonic core, thereby compromising hardware performance.
% Effectively, there exists a trade-off between energy efficiency and accuracy when designing analog computing cores. 
While the accuracy degradation in inference can be recovered for some models through sophisticated training methods such as quantization-aware training~\cite{quantization-wu-2020, quantization-krishnamoorthi-2018}, the impact of the abovementioned information loss on training accuracy is mostly irreparable.
% }

% \begin{figure}
%     \centering
%     \includegraphics[width=\linewidth]{figures/acc-tile-size.png}
%     \caption{Accuracy vs. vector size in conventional analog hardware for ResNet50 on ImageNet}
%     \label{fig:acc-tile-size}
% \end{figure}

\begin{figure*}[ht]
    \centering
    
    \includegraphics[width=\textwidth]{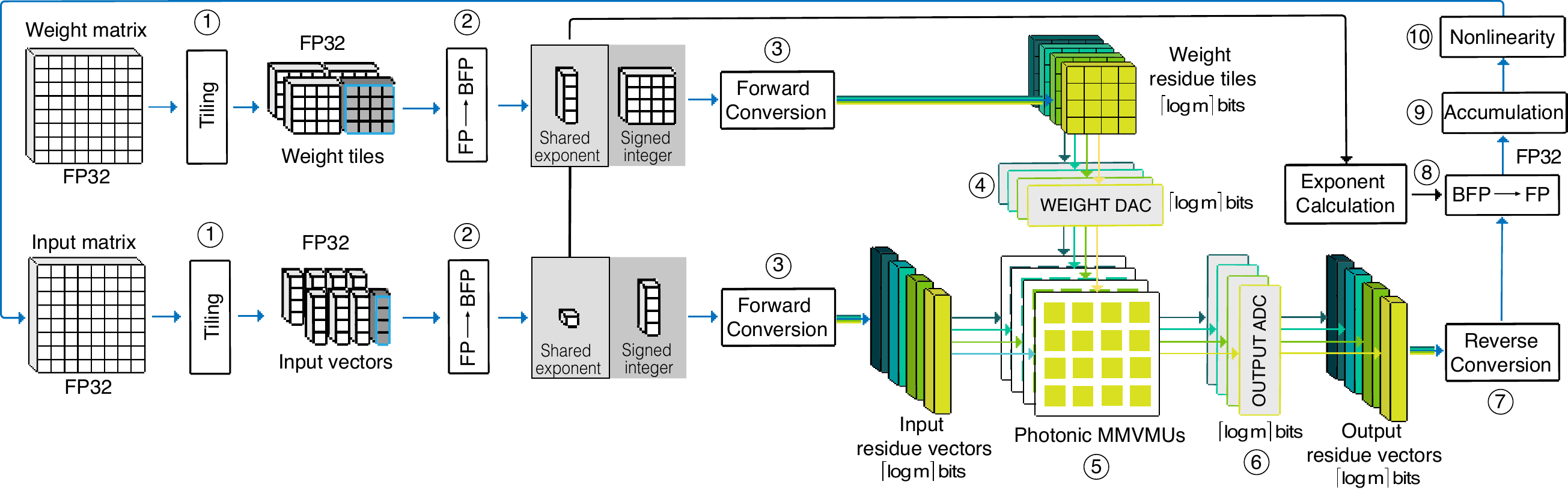}
    % \vspace{-0.15in}
    \caption{Mirage's RNS-based dataflow for a single tiled-MVM operation as part of a forward pass. We show a four-moduli case in this figure as an example.
    }
    % \vspace{-0.2in}
    \label{fig:dataflow}
\end{figure*}

\subsection{The Residue Number System (RNS)}
\label{sec:rns}
The RNS represents an integer as a set of smaller integers called residues, which are obtained by performing a modulo operation on the said integer using a selected set of $n$ co-prime moduli. 
As an example, consider an integer $X$. 
In RNS, $X$ is represented with $n$ residues $x {=} \{x_1, \text{...}, x_n\} $ for a set of $n$ moduli $\mathcal{M} {=} \{m_1, \text{...}, m_{n}\}$ where $x_i {=} |X|_{m_i} {\equiv} {X\bmod m_i}$ for $i {\in} \{1, \text{...}, n\}$. 
$X$ can be uniquely reconstructed from the residues and the corresponding moduli using the Chinese Remainder Theorem (CRT):
% \vspace{-0.1in}
\begin{equation}
X = \sum_{i=1}^n(x_i M_i T_i)_M,
    \label{eq:crt}
    % \vspace{-0.05in}
\end{equation}
if all the moduli are co-prime and $X\in [0, M)$ where $M{=} \prod_i m_i$. Here, $M_i {=} M/m_i$ and $T_i$ is the multiplicative inverse of $M_i$ (i.e., $|M_i T_i|_{m_i} {\equiv} 1$). 

The RNS is closed under addition and multiplication. 
Therefore, we can perform the GEMM operations in DNNs in the RNS space as long as we guarantee the output of the dot products stays within the RNS range (i.e., $[0, M)$).  
This $[0, M)$ range can be shifted to be symmetric around zero, i.e., $[-\psi,\psi]$, where $\psi {=} \lfloor (M{-}1)/2\rfloor$, to represent negative values.  
%{\color{red}
\subsection{Device Metrics and Noise Sources in Silicon Photonics}

\subsubsection{Modulation Mechanisms and Device Tradeoffs}
\label{sec:siph}
In silicon photonics, efficient reprogrammability is a critical property for devices such as MZIs and MRRs---which are the building blocks of the computing units. 
Modulation in such devices can be obtained through different mechanisms creating different tradeoffs between device metrics such as modulation bandwidth, optical loss, and device size---which impacts the scalability of the design. 
These mechanisms can broadly be grouped into three: thermo-optic, free-carrier-dispersion-based, and nano/micro-opto-electro-mechanical systems (N/MOEMS)-based devices. 
Thermo-optic devices are widely used in silicon photonics due to their simple fabrication process and high modulation efficiency, but their modulation speed is typically limited to a few KHz. 
In addition, the heaters used for modulation dissipate significant power and can easily cause thermal crosstalk.
The free-carrier-dispersion effect can be used to design high-speed modulators.
While these modulators can easily reach tens of GHz of bandwidth, they typically are lossy and require longer device lengths.
Lastly, N/MOEMS-based devices have recently emerged as a viable alternative to the other two mechanisms.
These devices~\cite{Baghdadi:21, ramey20, Feng:20noems} provide a moderate modulation frequency (up to a few hundred MHz) and low optical loss with negligible static power consumption. 
%It is crucial to carefully evaluate these methods and their tradeoffs between speed, energy, and area for achieving a feasible design in photonic systems. 
% }
% {\color{teal}
\subsubsection{{Sources of Analog Noise}}
\label{sec:noise}
Shot noise and thermal noise are the two main sources of noise in analog photonic cores~\cite{noise9932877}. 
Shot noise occurs due to the statistical variation in the number of photons or electrons. It can be approximately represented by a Gaussian distribution as
% \vspace{-0.05in}
\begin{equation}
I_S = \mathcal{N} \left( 0,\ 2{q_e}{I_{{\text{D}}}}\Delta f \right),
\end{equation}
% \vspace{-0.03in}
where $q_e$ is the elementary charge, $I_D$ is the photodetector current, and $\Delta f$ is the bandwidth. 
Thermal noise results from the resistor in the trans-impedance amplifier (TIA) circuitry, which can be modeled as
\begin{equation}
I_T = \mathcal{N}\left( 0,\ \frac{{4{k_B}T}}{{{R}}}\Delta f \right),
\end{equation}
where $k_B$ is the Boltzman constant, $T$ is the temperature, and $R$ is the feedback resistor of the TIA.
In the presence of noise, to achieve a desired bit precision $b$, one should be able to reach $2^b$ separable levels, i.e., SNR $\geq 2^b$.
It is a common practice to increase the optical input power to suppress the noise until the SNR is large enough to achieve $b$ bits~\cite{shiflett2021albireo, demirkiran2023electro}.
% }

% Prior works show that through careful device parameter choices, calibration, and error correction methods, these errors can be minimized for MZI- and MRR-based structures~\cite{sunny2021crosslight, 9634115mrr, Bandyopadhyay:21, Miller:15}.

% }

% In this section, we first present the RNS-based computing model and the dataflow in \ACC. This is followed by a description of the design of the photonic modulo-arithmetic units and the design of the \ACC \microarchitecture.
\section{RNS-Based Dataflow in \ACC}
\label{sec:mirage-dataflow}
RNS is closed under addition and multiplication allowing a GEMM operation to be performed in the RNS space. 
Using RNS, Eq.~\eqref{eq:nn} can be rewritten as:
% \vspace{-0.1in}
\begin{equation}
    \vec{X}^{(\ell+1)} = f^{(\ell)} \Bigg(\text{CRT}\bigg(\Big|
    \big|W^{(\ell)}\big|_{\mathcal{M}}
    \big|{X}^{(\ell)}\big|_{\mathcal{M}}
     \Big|_{\mathcal{M}}\bigg)\Bigg),
    \label{eq:nn-rns}
\end{equation}
where $\mathcal{M}$ represents a selected set of $n$ co-prime moduli. 
In the RNS space, each matrix is represented by $n$ residue matrices for $n$ moduli. 
GEMM in the RNS space is then a set of modular GEMM operations---one GEMM per modulus, $n$ GEMMs in total.
After the GEMM operations are performed, the resulting $n$ output residue matrices are converted back to a single output matrix in BNS by using Eq~\eqref{eq:crt}.
The same approach applies to the GEMM operations in the backward pass as stated in Eqs.~\eqref{eq:back-x} and~\eqref{eq:back-w}. 

Fig.~\ref{fig:dataflow} shows the dataflow for a tiled-GEMM operation as part of the forward pass in \ACC. 
The dataflow follows these steps:
% \begin{enumerate}

    \circled{1} The FP32 input and FP32 weight matrices (flattened if necessary) are tiled. 
    Tile size is equal to the size of the MMVMUs.
    
    \circled{2} The FP32 values are converted into BFP.
    In an MVM operation with BFP values, the input vector and each row of the weight tile represent a group\footnote{The group size in BFP, denoted as $g$, the vector size, denoted as $h$, and the horizontal dimension of the photonic MMVMU (the number of optical MAC units within a row), all signify the count of elements within the vectors subjected to a dot product. From this point onward, we will collectively refer to these three terms as $g$.}. 
    For each group, the largest exponent among the group elements is chosen to be the shared exponent ($e_{\vec v}$). 
    Given an $e_{\vec v}$ for a group $\vec v$ with a group size $g$, i.e., $e_{\vec v} = \text{max}(e_{{\vec v[i]}})\ \forall{i} \in \{1, \text{...},\  g\}$, the mantissae of the group elements are shifted right by the difference between the shared exponent and their original exponent, i.e., $m_{\vec v[i]}=m_{\vec v[i]}>> (e_{\vec v}-e_{{\vec v[i]}})$, where ${\vec v[i]}$ is the $i^\text{th}$ element of $\vec v$. 
   The LSBs of the mantissae are then truncated depending on the number of mantissa bits ($b_m$).
   
    \circled{3} We perform `forward conversion' to convert the `$b_m{+}1$'-bit signed integers, i.e., sign and mantissa bits of the BFP values, of the input vector and weight tile into the RNS space.
%    In the RNS space, each integer is represented by $n$ residues for $n$ moduli. 
    Forward conversion generates $n$ input vectors and $n$ weight tiles in total for $n$ moduli.
    %These $n$ input vectors and $n$ weight tiles contain the residues calculated during forward conversion.
    Here each residue is represented by $\lceil \text{log}_2(m_i)\rceil$ bits where $m_i$ is the $i^\text{th}$ modulus value. 
    Forward conversion is a modulo operation that can be simplified into a simple shift operation when special moduli sets are used (See Section~\ref{sec:moduli-selection} for details).
    
    \circled{4} Each weight residue tile is passed through $\lceil \text{log}_2(m_i)\rceil$-bit DACs to ensure no information loss and programmed into the MMVMUs.
    The input vectors do not require DACs thanks to our photonic core design as each individual digit of the binary input is multiplied separately (more details in Section~\ref{sec:modular-arit-ph}).
    
   \circled{5} Analog modular MVM operations are performed in the MMVMUs using cascaded phase shifter blocks. 
    The operations are inherently modular as the operands are encoded directly in the amount of phase shift applied to an optical signal (See Section~\ref{sec:modular-arit-ph}).
    
    \circled{6} The outputs are detected by photodetectors and converted into the digital domain by using $\lceil \text{log}_2(m_i)\rceil$-bit ADCs.
    Here, it is important to note that weight DACs and output ADCs have the same precision. 
    This is because the use of modular arithmetic in the analog domain ensures that the data bit-width does not grow during operations.
    Therefore, the output can be collected with no information loss at the ADCs with the same bit-width as DACs.
    
    \circled{7} The collected output residues are converted back to BNS.
    This `reverse conversion' can be implemented using Eq.~\eqref{eq:crt}. 
    %CRT, Mixed-Radix Conversion (MRC), or look-up tables. 
    Similar to forward conversion, using special moduli sets alleviates the complexity and overhead of this step.
    
    \circled{8} The exponents of the output vector are digitally calculated in parallel with the analog modular MVM operations. 
    Using the output mantissae and exponents, FP32 values are constructed.
    
    \circled{9} The partial outputs are accumulated to compose the final GEMM output. 
    The dataflow in \ACC requires the partial outputs to be written to the on-chip memory.
    For each partial output, a read-accumulate-write operation is performed.
    
    \circled{10} Steps 2-9 are repeated for each tile in a layer and nonlinearity is then applied to the final GEMM result (digitally in FP32). 

Steps 1-10 are repeated for each layer until the forward pass is complete. 
Input and weight gradients are then calculated in a similar manner where input and weight matrices in the diagram are replaced by the operands in Eqs.~\eqref{eq:back-x} and \eqref{eq:back-w}.
Once the weight gradients are obtained, the weight values are updated according to Eq.~\eqref{eq:w-update}. 
Here, we perform all the GEMM operations in BFP, however, we store the weights in FP32 in the memory and perform the weight updates in FP32. 

% It should be noted that Fig.~\ref{fig:dataflow} shows a weight-stationary dataflow where a weight tile is programmed into the photonic array and kept stationary while input vectors are updated each cycle while reusing the same weight tile. 
% An input-stationary dataflow can easily be implemented by swapping the input and weight matrices in the diagram.
% The output-stationary dataflow is not an efficient solution for \ACC as it requires reprogramming the optical devices in the photonic arrays (which is expensive in terms of either energy or time) every cycle. 
% The impact of different dataflow options on performance is discussed in detail in Section~\ref{sec:dataflow-eval}.

% To prevent overflow and guarantee the integrity of these tiled-MVM operations in the RNS space, we need to ensure that
% \begin{equation}
%     \log_2 M \geq b_\text{out} = b_\text{in} + b_w + \log_2(h)-1,
%     \label{eq:rns_bit_inequality}
% \end{equation}
% for $b_\text{in}$-bit inputs, $b_w$-bit weights, and $h\times h$ MVM tiles. 
% Here $b_\text{in}$ and $b_w$ refer to the bit-width of the signed integers in the BFP representation, i.e., $b_m+1$ where $b_m$ is the number of mantissa bits. 
% Eq.~\eqref{eq:rns_bit_inequality} requires a careful selection of the moduli set and BFP configuration (i.e., number of exponent and mantissa bits and group size). The impact of this selection process will be investigated later in Section~\ref{sec:moduli-selection}.

\section{{\ACC} \microarchitecture}
\label{sec:architecture}

In this section, we present the micro-architecture of \ACC enabling the RNS-based dataflow explained in Section~\ref{sec:mirage-dataflow}. 
We first describe the design of our proposed photonic modular arithmetic units, discuss the moduli selection process, and lastly, explain the accelerator design and how different components in \ACC interact.  

\subsection{Photonic Modular Arithmetic Units}
\label{sec:modular-arit-ph}

In \ACC, we perform tiled-MVM operations in the RNS domain, as shown in Fig.~\ref{fig:dataflow}, step 5. 
This step requires a new photonic core design for performing \emph{modular} arithmetic in the analog domain, unlike the conventional photonic cores relying on standard FXP arithmetic.
This section describes our novel modular MAC unit and how we build modular dot products and MVMs using this unit. 

\subsubsection{Modular Multiplication Unit (MMU)}
\label{sec:mmu}

\begin{figure}[t]
    \centering
    \includegraphics[width=\linewidth]{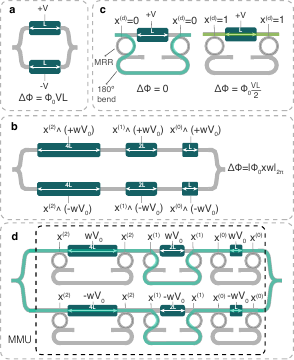}
    
    % \vspace{-0.05in}
    \caption{(a) Simple MZM with phase shifters with length $L$ and applied voltage $V$. (b) 3-bit modular multiplication using cascaded phase shifters. (c) Routing light using MRR switches. (d) 3-bit modular multiplication using MRR switches.
    % \vspace{-0.25in}
    }
    \label{fig:mmu}
\end{figure}

In a typical dual-rail Mach-Zehnder Modulator (MZM) (see Fig.~\ref{fig:mmu}(a)), the phase difference between input and output is
% \vspace{-0.1in}
\begin{equation}
    \Delta \Phi =  \frac{V L }{V_{\pi\cdot \text{L}}},
\end{equation}
where $V_{\pi\cdot \text{L}}$ is the phase shifter's modulation efficiency, which is a constant value. 
$\Delta \Phi$ is then proportional to both the length of the phase shifter, $L$, and the applied voltage, $V$. 
A regular multiplication, i.e., $xw$, is performed through \emph{amplitude} modulation using the attenuation caused by the phase shift on the input signal. 
However, in RNS, a modular multiplication $|xw|_m$ is needed. 
In \ACC, we obtain this behavior via \emph{phase} modulation. 
By encoding $x$ and $w$ in $L$ and $V$, we can obtain a phase shift that is proportional to $xw$ and inherently modular with $2\pi$ in the same MZM design, i.e., $\Delta \Phi \propto |xw|_{2\pi}$.

However, $L$ cannot be changed at runtime. 
Therefore, we use separate phase shifters for each digit of the binary operand, where the length of the shifter is proportional to the weight of the binary digit as shown in Fig.~\ref{fig:mmu}(b).
To encode a $b$-bit value, we use $b$ phase shifters with lengths $2^0L,\ 2^1L,\ \text{...},\ 2^{b-1}L$ for each bit from LSB to MSB and control each digit separately. 
For performing a multiplication, we map one operand ($w$ in this example) to the applied voltage and apply the same voltage to all $b$ digits. 
We then use the second operand ($x$) digit-by-digit to turn ON or OFF the voltage on each shifter separately.
This mapping requires an AND operation between the first operand and each digit of the second operand,
i.e., $V^{(d)} {=} (w V_0)\land x^{(d)}, \ \forall d\in\{0,\ ...,\ b-1\}$.

Fig.~\ref{fig:mmu}(b) shows a multiplication between two 3-bit integers where $w$ is encoded in the applied voltage as an analog value while the digits of $x$ control if $V$ is applied to each digit or not.
%Assume $V_0$ is the unit voltage and $\Phi_0$ is the unit phase shift obtained when $V_0$ voltage is applied to an $L$-long phase shifter.
For example, assume $x{=}101$ and $w{=}011$.
In this case, we set $V{=} wV_0{=}3V_0$ for all three digits\footnote{$V_0$ represents a unit voltage value that results in a unit phase shift ($\Phi_0$) in a $L$-long phase shifter. $\Phi_0$ is set to be $2\pi / m$ to perform a modulo $m$ operation, which is explained later in the section. 
For a given $\Phi_0$, the absolute values for $V_0$ and $L$ depend on the $V_{\pi\cdot \text{L}}$ of the phase shifters and the maximum available bias voltage.
}. 
As the LSB of $x$, $x^{(0)}{=}1$ and $1 {\land} V {=} 3V_0$, 
$3V_0$ is applied to the $L$-long phase shifter. 
This creates a $3\Phi_0$ phase shift on the optical signal passing through.
The second digit of $x$, $x^{(1)} {=} 0$. 
As $0 {\land} V{=} 0$, no voltage is applied to the $2L$-long phase shifter resulting in no phase shift on the optical signal.
Similar to the LSB, $x^{(2)}{=}1$ and $1 {\land} V {=} 3V_0$ voltage is applied to the $4L$-long phase shifter. 
This causes a $12\Phi_0$ phase shift as the phase shift is proportional to $V\cdot L$.
As the same optical signal goes through all the cascaded phase shifters, the sequentially introduced phase shifts are accumulated. 
By applying opposite signed voltages to the symmetrical arms of the dual-rail setup, a total of $(3{+}0{+}12)\Phi_0 = 15\Phi_0 \propto xw$ is applied to the signal ($15/2\ \Phi_0$ from each arm). 
The observed phase shift, however, is $|15\Phi_0|_{2\pi}$ as the optical phase is modular with $2\pi$.
\begin{figure*}[ht]
    \centering
    \includegraphics[width=\textwidth]{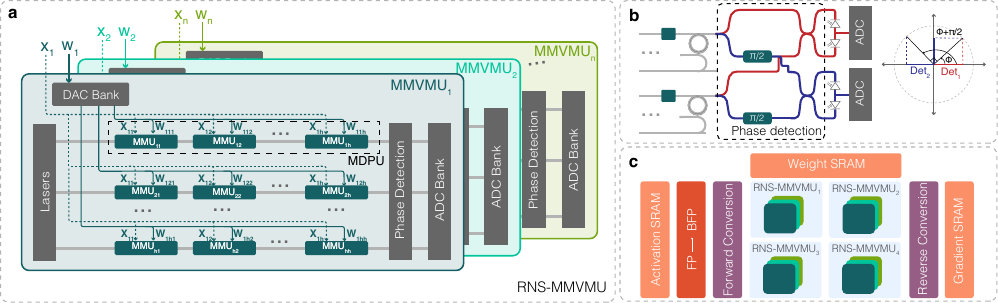}
    % \vspace{-0.1in}
    \caption{(a) RNS-based MMVM Unit (RNS-MMVMU) \microarchitecture. (b) Phase detection unit. The top arms of the two rows detect the amplitude of the incoming signals directly while the bottom arms apply $\pi/2$ radians phase shift and detect the amplitude. Phase detection is done by using these two amplitude values. (c) Main components of \ACC architecture with four RNS-MMVMUs and three moduli as an example. 
    % \vspace{-0.25in}
    }
    \label{fig:rns-mvm}
\end{figure*}
By adjusting $\Phi_0$ to be $2\pi/m$, modular arithmetic with arbitrary modulus $m$ instead of $2\pi$, i.e.,  $|xw|_m$, can be obtained as
% \vspace{-0.05in}
\begin{equation}
    \Delta \Phi_{\text{total}} = \big|\big(\sum_{d=0}^{b-1} (2^d x^{(d)}w\frac{2\pi}{m})\big)\big|_{2\pi} = \frac{2\pi}{m} \big| (xw)\big|_{m}.
    % \vspace{-0.05in}
\end{equation}
This adjustment is done through the unit applied voltage, i.e., $V_0 = 2 V_{\pi}/m$.
The resulting output value ($\Delta \Phi_\text{total}$) read at the end of the optical path is then multiplied back by $m/2\pi$ to obtain the output.
% }
% {\color{magenta}
%It should be noted that the MMU design with multiple phase shifters results in an increase in length compared to a typical MZM. 

For a modulus $m$, both $x$ and $w$ are both integer residues varying between $[0,\ m-1]$, which can be mapped around zero as $[-\lfloor \frac{m-1}{2} \rfloor , \lceil \frac {m-1}{2} \rceil ]$.
In this case, the maximum output on an MMU is $xw = \lceil \frac {(m-1)^2}{2} \rceil $, requiring a $\Delta \Phi_{\text{max}} = \lceil \frac{(m-1)^2}{2}\rceil \frac{2\pi}{m}$ phase shift. 
The MMU should be able to reach $\Delta \Phi_{\text{max}} $ when the bias voltage ($V_\text{bias}$) is fully applied. 
This requires a total phase shifter length of 
\begin{equation}
    L_\text{total} = \frac{V_{\pi L}}{V_\text{bias}}\frac{\Delta \Phi_{\text{max}}}{\pi}.
\label{eq:ps_len}
% \vspace{-0.1in}
\end{equation}

Given a $\Delta \Phi_{\text{max}} $, there is a trade-off between $V_\text{bias}$ and $L_\text{total}$ caused by the constant ${V_{\pi L}}$ of the phase shifters. ${V_{\pi L}}$ depends on the chosen actuation mechanism of the devices as mentioned in Section~\ref{sec:siph}.

% It should be noted that $L_\text{total}$ can be longer than a typical MZM as the required phase shift is larger.
%that is approximately proportional with $m$.
% }

% {\color{red} To make this approach feasible, we should carefully study device characteristics and tradeoffs in phase shifters. 
% Broadly, there exist three main phase-shifting mechanisms in silicon photonics: thermo-optic, MEMS-based, and free-carrier-dispersion-based. 
% The thermo-optic phase shifters are widely used for their low optical loss, however, they can only be reprogrammed at KHz bandwidths. In addition, thermal tuning causes thermal crosstalk between optical devices---which requires longer distances between devices or additional tuning circuits. 
% MEMS-based devices 
% }

In \ACC, the input operands of the MMU ($x$ and $w$) are integer elements of the residue matrices to be multiplied during a tiled-GEMM operation---which is a series of tiled-MVM operations. 
During the tiled-GEMM operation, one MVM is performed every cycle. For each MVM operation, at least one of the input operands needs to be updated. 
In the MMU design shown in Fig.~\ref{fig:mmu}(b), an update in $x$, $w$, or both, all result in reprogramming the phase shifters \emph{every cycle}.
In this case, phase shifters must have high modulation bandwidths ($\geq$GHz) to perform high-speed MAC operations. 
As mentioned in Section~\ref{sec:siph}, this high bandwidth in phase shifters can be obtained through free-charge dispersion, however, such devices are typically quite lossy and have relatively high $V_{\pi L}$ values (the lower the better) causing longer device lengths. 
Using these high-bandwidth phase shifters, one can achieve high-speed operations, but this can easily lead to ${\geq} 10$ dB optical loss and tens of mm length per MMU, significantly limiting the scalability of the design. 
Alternatively, one can use thermo-optic or M/NOEMS-based phase shifters with lower optical loss and lower $V_{\pi L}$. 
Yet, these tuning mechanisms require long delays ($\mu$s-to-ms) for reprogramming, limiting the clock speed of the photonic core to KHz to a few hundred MHz. 
Effectively, both options, high optical loss/longer device length or low modulation bandwidth, result in poor performance. 

To resolve this issue, we modified our design to leverage data stationarity during operations by encoding the two operands  ($x$ and $w$) onto separate devices, which is shown in Fig.~\ref{fig:mmu}(c) for a single MMU digit.
Here, instead of turning the applied voltage ON or OFF via a digital AND operation, we obtain the same behavior using MRR switches to change the route of the light to go through or bypass the phase shifter to avoid frequent reprogramming in phase shifters. 

MRRs are optical devices that are designed to have a resonant wavelength. 
If the wavelength of the signal on a waveguide that is next to an MRR matches the resonant wavelength of the MRR, the signal is coupled into the MRR, otherwise, it keeps propagating down the waveguide.
By applying a voltage to an MRR, its resonant wavelength, and therefore, the route of the light on the waveguide can be changed.
In Fig.~\ref{fig:mmu}(c), assume the default resonant wavelength (when no voltage is applied to the MRR) and the passed optical signal's wavelength are the same. 
If the corresponding binary digit of the input operand is zero, i.e., $x^{(d)}{=}0$, no voltage is applied to the MRR so the optical signal is coupled into the MRR and routed through the bypass waveguide with no phase shifter (Fig.~\ref{fig:mmu}(c), left).
In contrast, if $x^{(d)}{=}1$, a voltage high enough to shift the resonant wavelength is applied to the MRRs so that the input signal is not affected and it propagates on the upper arm containing the phase shifter (Fig.~\ref{fig:mmu}(c), right). 
Fig.~\ref{fig:mmu}(d) shows the same multiplication unit as Fig.~\ref{fig:mmu}(b) with MRR switches for the abovementioned example where $x=101$ and $w=011$.

In the modified design (Fig.~\ref{fig:mmu}(c-d)), as $x$ and $w$ are encoded onto separate devices, the change in $x$ does not impact the voltage applied to the phase shifter. 
Therefore, with a dataflow where $w$ is stationary, the voltage applied to the phase shifters ($wV_0$) can be kept fixed during MVM operations and requires reprogramming only once for each tiled-GEMM operation, instead of every cycle. 
By minimizing the number of times we have to reprogram the phase shifters, we can use more efficient and low-bandwidth phase shifters without compromising performance. 
The new design requires adding MRR switches, which introduces extra optical loss and increases the area. 
% However, MRR switches have been extensively used as high-speed switches in silicon photonics. 
However, previous works show that MRRs can easily achieve tens of Gb/s with a much smaller optical loss and area footprint compared to high-bandwidth phase shifters~\cite{Ohno:21}. 
Therefore, using a combination of MRR switches and low-bandwidth phase shifters
in the MMU enables us to achieve a high-speed and scalable design.

% {\color{red}The digits of $x$ are fed to the MRRs and used to change the route of light every cycle. 
% In this way, once $w$ is encoded in the voltage applied to the phase shifters, it can be kept stationary and reused for multiple $x$ values. 

% This enables us to use low-bandwidth low-loss phase shifters~\cite{Baghdadi:21} and leverage data reuse while achieving high speed.}

% To resolve this issue, instead of changing the applied voltage every cycle, we modified our MMU design to keep the values in the phase shifters ($w$) stationary and bypass the phase shifters when the single-bit digit of the operand is zero ($x^{(i)} {=} 0$). 
% To enable bypassing of phase shifters, we added MRR switches before and after each phase shifter and bypass waveguides to the unit. 

% MRRs change the route of the light and turn it in the opposite direction. 
% $180\degree$ bend waveguides are used to fix the direction of the light after passing through an MRR.  

\subsubsection{Modular Dot Product Unit (MDPU) and Modular MVM Unit (MMVMU)}
\label{sec:mdpu}
Similar to cascading phase shifters in a single MMU, we can cascade multiple MMUs to construct an MDPU and perform a modular dot product.  
The phase shifts introduced by each MMU accumulate and the modular dot product is obtained by measuring the total phase shift ($\Delta \Phi_{\text{total}}$) on the optical signal.
As the MMU operands are already scaled by $2\pi/m$, the dot product result will also be modular with $m$.
$\Delta \Phi_{\text{total}}$ in an MDPU with $g$ MMUs in a row is
\begin{equation}
\vspace{-0.1in}
    \Delta \Phi_{\text{total}} =\big| \sum_{j=0}^{g-1}\big(\sum_{i=0}^{b-1} (2^i x_j^{(i)}w_j \frac{2\pi}{m})\big)\big|_{2\pi} = \frac{2\pi}{m} \big| \sum_{j=0}^{g-1}(x_jw_j)\big|_{m}.
% \vspace{-0.1in}
\end{equation}
The final result of the dot product is collected at the end of the MDPU by detecting the optical phase and multiplying it by $m/2\pi$.

Multiple MDPUs construct an MMVMU and can perform a modular MVM operation in a single cycle.
In {\ACC}, we utilize $n$ MMVMUs to perform $n$ modular MVMs in parallel for an RNS with $n$ moduli.
The $n$ MMVMUs together form an RNS-MMVMU.
As illustrated in Fig.~\ref{fig:rns-mvm}(a), the input residue vector $x_i$ and weight residue tile $w_i$ for each modulus $i\in\{1,\ ...,\ n\}$ are sent to the $i^\text{th}$ MMVMU.
$x_i$ is broadcasted to all MDPUs within an MMVMU.
As $w$ is mapped to an analog value representing an integer smaller than $m$, it passes through $\lceil \text{log}_2m\rceil$-bit DACs without any information loss.
In contrast, $x$ is encoded digit-by-digit so the digits can directly be used to control the MRRs without DACs. 
The results from each MMVMU are detected via phase detection units and are passed through $\lceil \text{log}_2m\rceil$-bit ADCs.
These values are then converted to BNS via the reverse conversion unit.

\subsubsection{Phase Detection Unit}
\label{sec:detect}

In \ACC, the output of a modular MVM operation is stored in the phase of the output signal from an MMVMU.
However, a photodetector can only detect the amplitude of a signal. 
To detect the phase successfully, we need two amplitude measurements with $90\degree$ separation~\cite{Taylor2009PhaseEM}. 
Fig.~\ref{fig:rns-mvm}(b) shows the detection setup in \ACC.
Here, the idea is to read both $x$ and $y$ coordinates in the polar plane to determine the phase angle. 
We first directly measure the amplitude ($x$ coordinate). 
Then, we apply a $\pi/2$ phase shift and measure the amplitude again ($y$ coordinate).
The combination of these two measurements is unique to the corresponding phase level.
To measure the $x$ and $y$ components separately, we split both arms in the dual-rail setup into two. 
The upper splits of both arms (output of the MDPUs) are directly sent to the first set of balanced photodetectors. 
We apply a $\pi/2$ phase shift to the lower splits of the two arms and send them to the second set of balanced photodetectors.
It should be noted that this setup requires two detections and two ADCs per MDPU and twice the laser power to be injected (compared to a single-detection setup). 

\subsection{Moduli Selection}
\label{sec:moduli-selection}
Moduli selection plays a crucial role in designing {\ACC}. 
For a chosen $\mathcal{M}$, there is a limited range of values that can be uniquely represented. 
This range is called the dynamic range of the RNS and is $[0, M)$, where $M{=}\prod_i m_i$. 
To preserve the integrity of operations, the residues in the RNS space must stay within the RNS range, limiting the bit-width of the operands and the number of operations that can be performed in the RNS space.
To this end, for a tiled-MVM operation in the RNS space, we need to ensure that
\begin{equation}
    \log_2 M \geq b_\text{out} = 2 (b_m {+}1)  + \log_2(g)-1,
    \label{eq:rns_bit_inequality}
\end{equation}
for a BFP configuration with $b_m$ bits of mantissa and a group size of $g$. 
Here, $b_m{+}1$ is used for $b_\text{in}$ and $b_w$ as both inputs and weights use the same BFP configuration. 

In \ACC, the DNN accuracy is determined by the chosen $b_m$ and $g$ and is independent of the exact values of the moduli as there is no information loss during RNS operations as long as $M$ is chosen to be large enough to guarantee we satisfy Eq.~\eqref{eq:rns_bit_inequality}.
However, the selected moduli set has a significant impact on the hardware performance. 
Higher $b_m$ or $g$ naturally dictates a larger $M$. 
A larger $M$ requires either a higher number of moduli or larger moduli values. 
While the number of moduli determines the number of MMVMUs in {\ACC}, the bit-width of the moduli determines the bit precision of the data converters and the SNR that needs to be maintained in the photonic core.
%Both of these, in turn, determine the efficiency of {\ACC}.

Importantly, the selection of moduli impacts the cost of the forward and reverse conversion circuits. 
Typically, as $M$ and the moduli values get larger, these conversions get slower and more energy-consuming. 
Several works showed that traditional conversion methods such as CRT pose performance limitations for high dynamic range when arbitrary moduli sets are used~\cite{Wang:2002}. 
In a high-speed low-energy design such as \ACC, these conversions can easily become the bottleneck.
Instead, using special moduli sets and conversion hardware can alleviate the hardware overhead of these operations significantly.
% {\color{magenta}
In \ACC, we use a three-moduli set in the form of $\{2^k{-}1,\ 2^k,\ 2^k{+}1\}$ where $k$ is a positive integer~\cite{hiasat2019residue}. 
This set reduces modulo operations into simple shift operations. 
During the forward conversion, $|A|_{2^k} {=} A {>>} k$, $|A|_{2^k{+}1} {=} |A|_{2^k} {+} 1$ (subtract $2^k+1$ if $\geq2^k+1$), and $|A|_{2^k{-}1} = |A|_{2^k} {-}1$ (add $2^k{-}1$ if ${<}0$). 
The reverse conversion is typically more costly than the forward conversion due to the modulo $M$ operation with large $M$ values. 
Similar to the forward conversion, the special moduli set can simplify this operation as $M =2^{3k}{-}2^{k}$ and $|R|_{M} = |R|_{2^{3k}} {-} |R|_{2^{k}} = (R{>>}3k) {-} (R{>>}k)$ (add $M$ if $<0$).
Previous works show that this design can provide a high dynamic range (up to 24 bits) with ${\sim}2$ GHz throughput with very low power consumption (${\sim}1$ mW). 
Please refer to Hiasat~\cite{hiasat2019residue} for implementation details.
% }

\subsection{\ACC Accelerator Design}
\label{sec:mirage-design}
% {\color{blue}
Fig.~\ref{fig:rns-mvm}(c) shows the main components of \ACC. 
\ACC consists of a photonic and an electronic chiplet that are integrated via 3D integration. 
When executing a DNN layer, first, the input and weight matrices are tiled and the FP-to-BFP and BNS-to-RNS (forward) conversions are performed on these tiles (steps 1-3 in Fig.~\ref{fig:dataflow}). 
These operations are handled by the electronic chiplet. 
The integer residues obtained by the forward conversion are then sent to the photonic chiplet (after passing through DACs if the data are mapped to analog voltages). 
Each input vector-weight tile residue pair for a modulus is sent to the corresponding MMVMU on the photonic chiplet (See Section~\ref{sec:modular-arit-ph} for details about MMVMUs).  
The outputs of the analog modular MVM operations are collected from the photonic chiplet through photodetectors and the TIA circuitry that are placed on the electronic chiplet. 
The results are converted back to the digital domain via ADCs. 
The RNS-to-BNS (reverse) conversion is then performed on the output residues and the values are converted back to FP from BFP on the electronic chiplet. 
The outputs of the tiled-GEMM operations are accumulated and nonlinearity (ReLU, MaxPool, etc.) is applied digitally in FP32.
Steps 7-10 are performed via dedicated electronic circuitry in \ACC. 
% }

The data is read/written from/to the on-chip SRAM arrays. 
In our design, there are three separate SRAM arrays for storing activations, weights, and gradients, that are placed on the electronic chiplet along with the other digital circuitry.
In \ACC, the photonic circuit is clocked at $10$ GHz, whereas the digital circuits are clocked at $1$ GHz.
It is crucial to match the throughput of the digital electronic components with the photonic compute unit as digital operations are much slower and can easily become the bottleneck in the accelerator.  
For this purpose, we use $10$ copies of each digital circuit that are interleaved by $0.1$ ns.
Each RNS-MMVMU has its own $10$ dedicated SRAM sub-arrays for each SRAM type.
Every $0.1$ ns cycle, an RNS-MMVMU reads and writes from/to one of these $10$ interleaved SRAM sub-arrays. 
The same approach is used for digital conversion circuits. 
For each RNS-MMVMU, there exist $10$ RNS-BNS converters and $10$ FP-BFP converters, each triggered with $0.1$ ns offset. 
This interleaved structure enables memory accesses and digital compute to be fast enough such that it does not limit the performance of the photonic core even though the SRAM sub-arrays and digital circuits individually have a $1$ ns clock period~\cite{demirkiran2023electro}. 
All operations, i.e., SRAM reads, BFP conversions, forward conversions, DAC operations, modular MVMs, photodetections, ADC operations, reverse conversions, accumulations, and SRAM writes, are pipelined to achieve a 10 Giga MVM per second throughput in each RNS-MMVMU.

%The photonic chiplet includes phase shifters and MRRs while the electronic chiplet includes the digital circuitry for nonlinear operations, accumulators, FP-BFP and RNS-BNS conversions, SRAM arrays for storing data, DACs, ADCs, and detection units.
%The photonic devices are interfaced with the electronics via traditional transceiver circuits. 

%The DACs control the analog voltage to program the phase shifters. 
%The output signal is detected with a photodetector and the photocurrent is converted into voltage via a TIA circuitry before being sent to ADCs. 

\section{Evaluation Methodology}
\label{sec:evaluation_method}
\subsection{Accuracy Modeling}
\label{sec:eval_accuracy}
We modeled the accuracy of \ACC in PyTorch using customized GEMM layers. 
In all models, we swapped each GEMM operation, i.e., convolution and linear layers, with our customized BFP versions for a given $b_m$ and $g$. 
Once the values are converted to the BFP format, BNS-RNS and RNS-BNS conversions and the chosen moduli set have no direct impact on the DNN accuracy as long as the RNS operations are guaranteed to stay within the RNS range. 
So we omit these conversions in the accuracy model for faster training experiments. 
% {\color{teal}We assume that the impact of analog noise on accuracy can be avoided with a high enough SNR, which is taken into account in the power consumption estimation. Therefore the accuracy estimations }

In our customized GEMM layers, the tensors are first flattened and grouped. 
For each BFP group, we calculate the shared exponent and align the mantissae for a given $b_m$ and $g$. 
We then perform the convolution or linear operation and collect the result.
This BFP conversion is done for all GEMM layers during both forward and backward passes of each layer. 
While the gradients are calculated in BFP, we make the weight updates in FP32. 
For this purpose, we store the weights in FP32 instead of BFP and call them within the optimizer right before the parameter update step.
After updating the weights, we switch back to the BFP format before the next forward pass. 
% BNS-RNS and RNS-BNS conversions and the chosen moduli set have no impact on the accuracy as long as the operations are guaranteed to stay within the RNS range. 
% So we omit these conversions in the accuracy model for faster experiments. 

\subsection{Hardware Performance, Power and Area}
\label{sec:eval_ppa}
\subsubsection{Photonic Devices and Lasers}
\label{sec:eval_met_ph}
The latency of the photonic RNS-MMVMUs in \ACC for GEMM operations is calculated through an in-house simulator.
This simulator calculates the number of tiles and the number of operations per tile within a DNN layer given the hardware configuration and layer shapes. 
For each tile, the reprogramming of phase shifters (similar design to Baghdadi et al.~\cite{Baghdadi:21}, internally simulated) takes $5$ ns during which the photonic compute core is inoperable.
Once the values in the phase shifters are settled, one RNS-MMVM operation is completed every $0.1$ ns ($10$ Giga MVMs per second). 
This operation rate is based on the modulation bandwidth of the MRRs~\cite{Ohno:21}.
ADCs~\cite{adc-7573535} achieve ${\geq} 10 $ GS/s sampling rate so they do not cause a latency overhead when the operations are pipelined. 
%The SRAM arrays and all the digital circuits are clocked at $1$ GHz and ten of them are used together with $0.1$ ns offset to match the $10$ GHz frequency of the photonic core.

The photonic core power consumption includes the laser source power and MRR tuning power. 
% {\color{teal}
The laser power injected into the MMVMUs needs to ensure that a target SNR, which is dependent on the modulus value, is achieved. 
For a modulus $m$, we should be able to differentiate $m$ phase levels ($\text{log}_2m$ bits), i.e., SNR${>}m$ where the noise includes shot and thermal noise mentioned in Section~\ref{sec:noise}. 
From the photodetector, we back calculate the required laser power that can maintain an adequate SNR accounting for all the optical losses on the optical path.
% }
% {\color{red}
The phase shifters have a $V_{\pi L}$ = $0.002$ V$\cdot$cm modulation efficiency and $1.6$ dB/mm loss. 
% }
%The $25\ \mu$m actuation length of the phase shifter causes $0.04$ dB loss and can achieve $5.65\pi$ radian phase shift~\cite{Baghdadi:21}.}
% {\color{teal}
The tuning cost of the phase shifters is negligible (a few fJ/bit).
Each MRR has a radius of $ 10\ \mu$m and a total (insertion and propagation) loss of $0.2$ dB when coupled with the light~\cite{Ohno:21}. 
MRRs use electro-optical tuning and have a very small power consumption of $0.3$ pW for switching~\cite{Ohno:21}. 
This power dissipation is ${\sim}10^7\times$ smaller than thermo-optic shifters which resolves the thermal crosstalk problem in MRRs~\cite{Ohno:21}.
% }
Each $180\degree$ bend waveguide has a $5\ \mu$m radius and $0.01$ dB insertion loss~\cite{bahadori2019universal}. 
The laser-to-chip coupler has a $0.2$ dB loss~\cite{hu2023ultrabroadband} and the laser has a $20\%$ efficiency~\cite{mourou2013future}. 
% \revC{
The length of the phase shifters varies depending on the modulus value in the selected set $\{2^k{-}1, 2^k, 2^k{+}1\}$  where $k=5$ (choice of $k$ will be justified in Section~\ref{sec:sensitivity-analysis}). 
Using the Eq~\eqref{eq:ps_len} and device metrics ($V_{\pi L}$ = $0.002$ V$\cdot$cm  and $V_\text{bias}{=}1.08$V), the total phase shifter length for the largest moduli $33$ can be calculated as $0.57$ mm. 
With the MRRs included, the total horizontal length of a single MMU becomes $0.8$ mm. 
% For an MMVMU, first, the number of optical devices and the total area per MMVMU is calculated. 
% The area of all MMVMUs is added together to calculate the total area footprint of the photonic chiplet in \ACC.  
% }

\subsubsection{Digital Circuitry}
\label{sec:eval_met_dig}
% {\color{orange}
The output signal of the photonic core is converted to the electrical domain through photodetectors and TIAs. 
The photodetectors have a $1.1$ A/W responsivity.
The TIAs consume $57$ fJ/bit~\cite{Rakowski18}.
Each $6$-bit DAC with $20$ GS/s sample rate consumes $136$ mW power and takes up $0.072\ \text{mm}^2$ area~\cite{dac-8303762}. 
Each $6$-bit ADC with $24$ GS/s sample rate consumes $23$ mW power and takes up $0.03\ \text{mm}^2$ area~\cite{adc-7573535}.
As DACs in \ACC are used only once for each tile and ADCs perform conversions every $0.1$ ns ($10 {<} 24$ GS/s), the power consumption of DACs and ADCs is amortized over the total training time. 
The bit precision required for DACs and ADCs is determined by the moduli set $\{2^k{-}1, 2^k, 2^k{+}1\}$ where $k=5$ (choice of $k$ will be justified in Section~\ref{sec:sensitivity-analysis}).
For $m{=}2^k{+}1$ with $k{=}5$, $\lceil\text{log}_2m\rceil{=}6$ so we use the 6-bit DACs and ADCs as is. 
For $2^k$ and $2^k{-}1$, $\lceil\text{log}_2m\rceil{=}5$ so we scale the energy consumption down by $1$ bit~\cite{murmann21mixed}.
% }
All three SRAM arrays (activation, weights, and gradients) are generated using the SRAM compiler for TSMC $40$ nm technology node~\cite{tsmc}. 
In \ACC, we employ three SRAM arrays, each with $8$ MB size, consisting of $32$ kB memory banks with an access latency of ${\leq}1$ ns. 
The BFP-FP and BNS-RNS conversion circuits are implemented in RTL and synthesized using Cadence Genus~\cite{genus} and the TSMC 40 nm library with a clock rate of $1$ GHz. 
Each BFP-FP unit consumes $1.32$ pJ and has a $1318.4\mu\text{m}^2$ area footprint.
% \revC{
Each BNS-RNS unit consumes $0.17$ pJ with a $231.7\mu\text{m}^2$ area footprint
Each RNS-BNS unit consumes $0.48$ pJ energy per conversion and requires $1545.8\mu\text{m}^2$ area~\cite{hiasat2019residue}. 

For comparison, we use systolic arrays that support several data formats including FP32~\cite{hardfloat}, BFLOAT16~\cite{bfloat16}, HPF8~\cite{sun2019hybrid}, INT8, INT12, and FMAC~\cite{zhang2022fast}.
We chose systolic arrays for their common usage in DNN acceleration and their superior performance over CPUs and GPUs~\cite{Eyeriss2017, tpuv3}. 
We implemented MAC units with the abovementioned data formats in RTL except FMAC for which the energy number is obtained from the recent work by Zhang et al.~\cite{zhang2022fast}.
The power and area per MAC unit are collected through synthesis using Cadence Genus and the TSMC $40$ nm library. 
%The FP formats operate at $500$ MHz while INT formats operate at $1$ GHz clock. 

\section{Evaluation Results}
\label{sec:evaluation}

In this section, we first conduct a sensitivity analysis to investigate the impact of different design knobs in \ACC to decide the optimal BFP configuration and hardware parameters. 
We then provide accuracy and performance, power, and area results for \ACC and compare \ACC against traditional systolic arrays with MAC units based on different data formats. 
Lastly, for completeness, we compare the performance of \ACC running DNN inference against prior photonic and electronic DNN inference accelerators. 

\subsection{Sensitivity Analysis}
\label{sec:sensitivity-analysis}
In this section, we analyze the impact of the number of mantissa bits ($b_m$) and group size ($g$) in the BFP representation and the accuracy-energy consumption tradeoffs introduced by these choices. 
After selecting the optimal $b_m$ and $g$, we perform sensitivity analysis for the number of MDPUs in an MMVMU and the number of RNS-MMVMUs in \ACC. 
Lastly, we explore several dataflows for \ACC and the systolic array to improve utilization.

\subsubsection{BFP Parameters}
\begin{figure}[t]
    \centering
    \includegraphics[width=\linewidth]{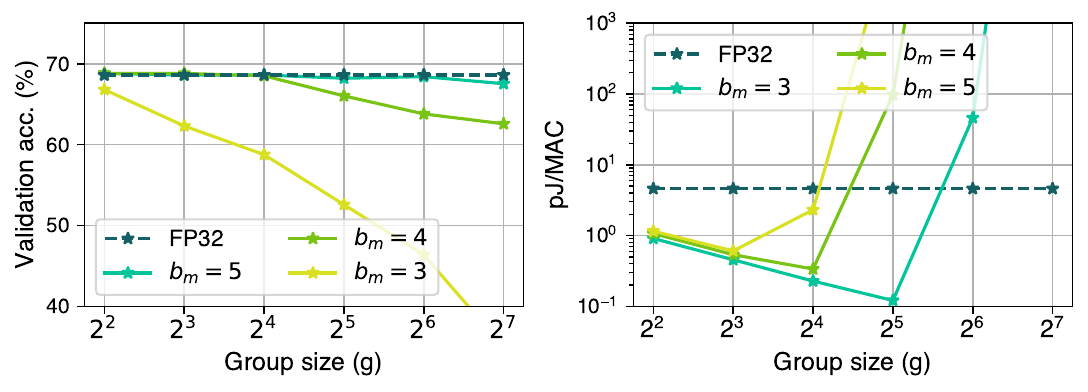}
    % \vspace{-0.15in}
    \caption{(a) ResNet18 validation accuracy on Imagenet after training from scratch for 60 epochs and (b) energy per MAC operation (pJ/MAC) for varying $b_m$ and $g$. 
    This analysis includes energy consumed by lasers and tuning circuitry, TIAs, DACs and ADCs, FP-BFP, and RNS-BNS conversions. Here, ResNet18 is shown as an example. We observed similar behavior for other evaluated DNNs. 
    % \vspace{-0.1in}
    }
    \label{fig:acc_sensitivity}
\end{figure}

The BFP configuration, i.e., $b_m$ and $g$ choice, significantly impacts the accuracy and hardware performance of \ACC. 
Fig.~\ref{fig:acc_sensitivity}(a) shows the validation accuracy after training ResNet18 with varying $b_m$ and $g$ in \ACC. 
The results indicate that we cannot reach high accuracy when $b_m{=}3$ and the minimum $b_m$ we can use is $4$ to achieve comparable accuracy to FP32 training.  
However, choosing $b_m{=}4$ allows us to go up to only $g{=}16$ without a drop in accuracy. 
When $b_m{=}5$, we can go up to higher $g$ values (up to $64$), which enables us to perform more MAC operations in parallel. 
This encourages us to take a deeper look into the $b_m{=}4$ and $b_m{=}5$ cases. 

Using the $\{2^k{-}1, 2^k, 2^k{+}1\}$ moduli set, the minimum $k$ we can choose that satisfies Eq.~\eqref{eq:rns_bit_inequality}, $k_\text{min}{=}4$ when $b_m{=}3$. 
Similarly, $k_\text{min}{=}5$ for $b_m{=}4$, and $k_\text{min}{=}6$ for $b_m{=}5$. 
In Fig.~\ref{fig:acc_sensitivity}(b), we compare the energy consumption per MAC operation of an RNS-MMVMU consisting of $3$ MMVMUs (one for each modulus) for varying $g$ and $b_m$.
This energy consumption includes lasers, MRR tuning, DACs and ADCs, TIAs, FP-BFP and RNS-BNS conversions. 
While a higher $g$ increases the number of MACs performed in parallel and helps amortize the cost of the components over $g$ MACs, it also requires more optical elements cascaded in an optical channel and increases the optical loss.
A higher optical loss requires an exponentially higher laser power in the photonic array to maintain the same SNR. 
As it can be seen from Fig.~\ref{fig:acc_sensitivity}(b), $b_m{=}4$ when $g{=}16$ provides the best energy efficiency among all the options that yield high accuracy in Fig.~\ref{fig:acc_sensitivity}(a).
Considering these results, in \ACC, we choose $b_m{=}4$ and $g{=}16$ and use these values for the rest of the experiments.

% To minimize the latency of these digital circuits, we considered these implementations for special moduli sets. We investigated the impact of moduli choice on energy consumption per MAC in the photonic core. Fig.~\ref{fig:energy_sensitivity} shows the energy consumption

% \begin{table}[h]
% \caption{}
% \label{tab:my-table}
% \begin{tabular}{|c|c|cc|cc|}
% \hline
%  &  & \multicolumn{2}{c|}{$b_m=4$} & \multicolumn{2}{c|}{$b_m=5$} \\ \hline
%  & Moduli set & \multicolumn{1}{c|}{$k_\text{min}$} & $g_\text{max}$ & \multicolumn{1}{c|}{$k_\text{min}$} & $g_\text{max}$ \\ \hline
% $S_1$ & $\{2^k{-}1, 2^k, 2^k{+}1\}$ & \multicolumn{1}{c|}{5} & 32 & \multicolumn{1}{c|}{6} & 64 \\
% $S_2$ & $\{2^k{-}1, 2^k, 2^k{+}1,  2^{2k}{+}1\}$ & \multicolumn{1}{c|}{3} & 32 & \multicolumn{1}{c|}{4} &  256\\ 
% $S_3$ & $\{2^k{-}1, 2^k, 2^k{+}1, 2^{k{-}1}{-}1\}$ & \multicolumn{1}{c|}{4} & 16 & \multicolumn{1}{c|}{5} &  64\\ 
% $S_4$ & $\{2^k{-}1, 2^k, 2^k{+}1, 2^{k{-}1}{-}1\}$ & \multicolumn{1}{c|}{4} & 128 & \multicolumn{1}{c|}{5} & 256 \\ 
% $S_5$ & $\{2^k{-}1, 2^k, 2^k{+}1, 2^{k{+}1}{-}1, 2^{k{-}1}{-}1\}$ & \multicolumn{1}{c|}{3} & 16 & \multicolumn{1}{c|}{4} & 128 \\ \hline
% \end{tabular}
% \end{table}

\subsubsection{RNS-MMVMU Array Size and Number of Arrays}

\begin{figure}[t]
    \centering
    \includegraphics[width=\linewidth]{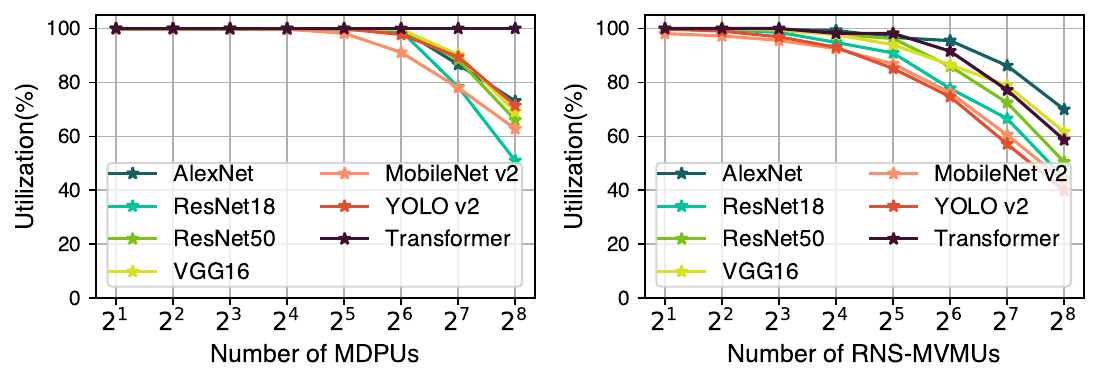}
    % \vspace{-0.15in}
    \caption{(a) Number of MDPUs versus spatial utilization ($\%$). (b) Number of RNS-MMVM units versus spatial utilization ($\%$).
    % \vspace{-0.3in}
    }
    \label{fig:utilization}
\end{figure}

As mentioned earlier, $g$ controls the horizontal array size of the RNS-MMVMU, i.e., the number of MMUs in each MDPU. 
We can, however, increase the vertical size by increasing the number of optical channels (MDPUs) for higher throughput. 
Additionally, we can utilize multiple RNS-MMVMUs on the same chip to further improve parallelism. 
Fig.~\ref{fig:utilization} (a) and (b) show the spatial utilization (how fully the MMVMUs are used) for a varying number of MDPUs in an MMVMU and a varying number of RNS-MMVMUs when $g{=}16$, respectively. 
The spatial utilization starts decreasing for almost all DNN models after $32$ MDPUs per MMVMU. 
%Therefore, we picked the number of MDPUs in each MMVMU in \ACC to be $32$.
In Fig.~\ref{fig:utilization} (b), we fix the MMVMU array size to be $16\times 32$ and increase the number of RNS-MMVMUs in \ACC. 
Here, we observe a decline in spatial utilization after $8$ RNS-MMVMUs for most models.
Considering these experiments, in \ACC, we choose MMVMU array size to be $16\times 32$ and the number of RNS-MMVMUs to be $8$.


\subsubsection{Dataflow Choice}
\label{sec:dataflow-eval}

Dataflow choice has been shown to have 
a critical impact on the performance of DNN hardware accelerators~\cite{Eyeriss2017}. 
For DNN inference, the typical dataflows can be listed as weight stationary, input stationary, and output stationary. 
The performance of these dataflows depends on the DNN model (layer shapes and sizes), the chosen batch size, and the underlying hardware. 
In this section, we investigate the impact of different dataflow options on the performance of \ACC and traditional systolic arrays. 
The dataflow names are intuitive for DNN inference. 
However, during training, we perform three GEMM operations per layer and the operands change for each operation. 
In the forward pass, we perform $O{=}WX$. In the backward pass, we perform $\Delta X{=}W^T\Delta O$ and $\Delta W{=}\Delta O X^T$.

To avoid confusion, we renamed these three dataflows, weight, input, and output stationary, to DF1, DF2, and DF3, respectively. 
DF1 is equivalent to the weight stationary dataflow where the first operands ($W$ for the forward pass, $W^T$ for $\Delta X$ calculation, $\Delta O$ for $\Delta W$ calculation) in the abovementioned GEMM operations are kept stationary, DF2 is equivalent to the input stationary dataflow where the second operands ($X$ for the forward pass, $\Delta O$ for $\Delta X$ calculation, $X^T$ for $\Delta W$ calculation) are kept stationary, and DF3 is equivalent to the output stationary dataflow where the output of the GEMM operations ($O$,  $\Delta X$, and $\Delta W$, respectively) are kept stationary. 
\begin{figure}[t]
    \centering
    \includegraphics[width=\linewidth]{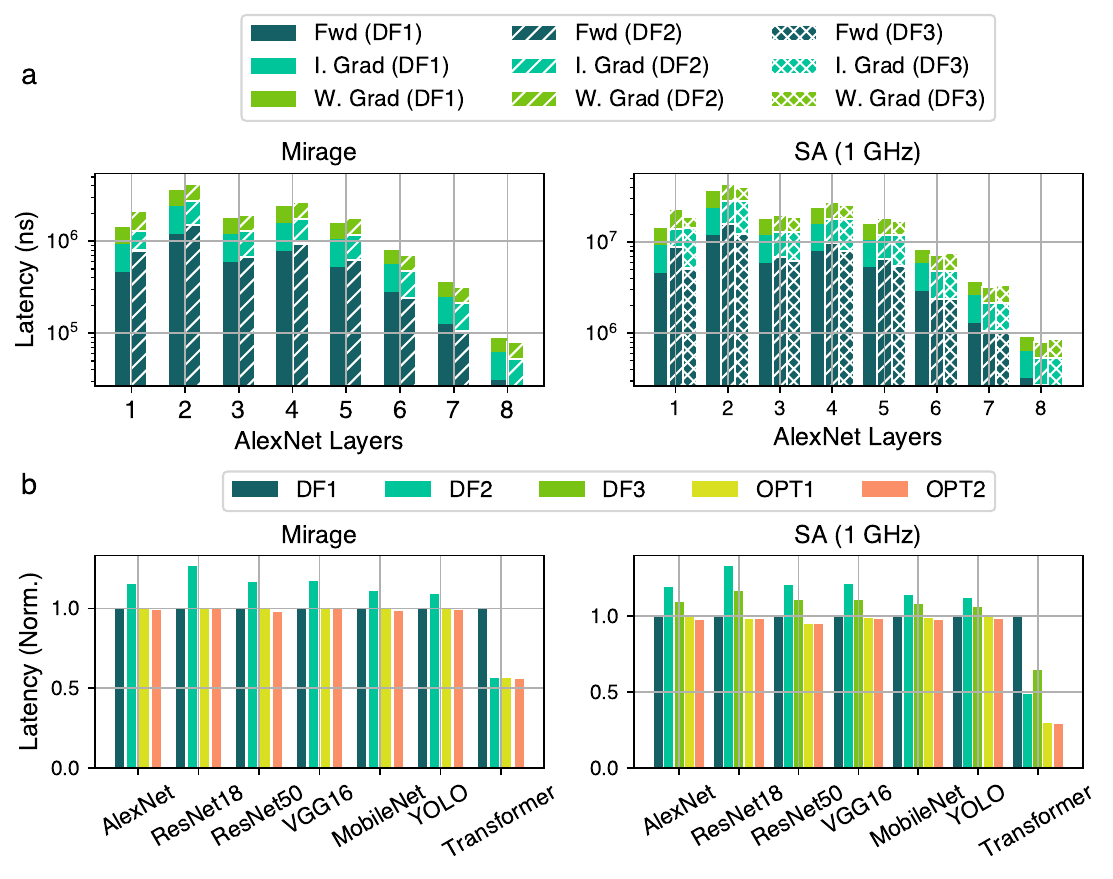}
    % \vspace{-0.1in}   
    \caption{(a) Latency per step for each layer of AlexNet for \ACC (left) and a $1$ GHz digital systolic array (right). (b) Latency per step for different DNNs and impact of dataflow for \ACC (left) and a $1$ GHz digital systolic array (right). The numbers for all dataflows are normalized to the DF1 results for all models. 
    %OPT1 and OPT2 show the flexible dataflow scenarios.
    % \vspace{-0.2in}
    }
    \label{fig:dataflow_opt}
    
\end{figure}

Fig.~\ref{fig:dataflow_opt}(a) shows the latency per training step (a single batch of 256) for different dataflows when running AlexNet on \ACC and a traditional systolic array with the same array size as \ACC and a clock frequency of $1$~GHz.  
In \ACC, we only consider DF1 and DF2 dataflows. 
This is because the DF3 dataflow requires both operands to be modified every cycle in the photonic arrays. 
However, as discussed in Section~\ref{sec:modular-arit-ph}, the modulation bandwidth of phase shifters is a limiting factor on the operation speed in the photonic core.
Therefore, it is preferable to minimize the number of updates in phase shifters to achieve high utilization of the photonic core.
DF1 encodes the first operand in the phase shifters and DF2 encodes the second operand in the phase shifters. 
In both cases, the values encoded in the phase shifters are kept stationary--which allows for high operation frequency. 
All three dataflows are applicable to systolic arrays. 

In Fig.~\ref{fig:dataflow_opt}(a), we observe that different dataflows perform better for different computations and different layers in a DNN model. 
For example, in the first layer of AlexNet, DF1 achieves a lower latency in the forward pass while DF2 achieves a lower latency in the input gradient calculation in \ACC.
A similar observation can be made for the systolic array design.
So to maximize performance, in both \ACC and the systolic array, we added flexibility in the choice of the dataflow. 
Fig.~\ref{fig:dataflow_opt}(b) shows the impact of using different dataflows as well as the added data flexibility optimizations (OPT1 and OPT2) for different DNNs. 
OPT1 chooses the best dataflow for each type of computation (i.e., forward pass, input gradient, and weight gradient calculation) which is kept the same for all layers. 
A more aggressive optimization, OPT2, picks the best dataflow for each GEMM operation separately for each layer. 
This dataflow scheduling is done once and offline for a DNN via analytical performance estimations. 

In Fig.~\ref{fig:dataflow_opt}(b), for \ACC, we observe that DF1 performs better for all models except Transformer in which DF2 achieves a better performance.
In all DNNs, the flexible dataflows (OPT1 and OPT2) bring minor to no benefit in \ACC.
For the systolic array, however, there exists more variety in the performance of different dataflows. 
On average across all reported models, OPT1 boosts the performance by $11.7\%$ and OPT2 boosts the performance by $12.5\%$ over the best-performing dataflow.
Although the OPT1 and OPT2 optimizations do not improve the performance of \ACC by much, we believe that it is important to consider this performance boost in systolic arrays to have the best possible baseline for comparison. 
For this purpose, we use OPT2 for both \ACC and systolic arrays for the rest of the analysis.

\subsection{Accuracy Evaluation}
\label{sec:accuracy-eval}
We evaluated \ACC's accuracy in commonly deployed CNNs for image classification on the ImageNet dataset~\cite{deng2009imagenet}, in YOLO-v2~\cite{redmon2017yolo9000} for object detection on the PASCAL VOC2012 dataset~\cite{everingham2010pascal}, and in a transformer~\cite{vaswani2017attention} model for machine translation on the IWSLT14 German-English dataset~\cite{bojar2014findings}. 
The CNN models were trained for $60$ epochs using the SGD optimizer with a batch size of $256$ and a learning rate starting from $0.01$ and scaled down by $10$ after each $20$ epoch. 
YOLO-v2 was trained for $120$ epochs using the SGD optimizer with an initial learning rate of $10^{-4}$, and scaled down by $10$ after epochs $60$ and $90$. 
The 12-layer transformer model with $12$ heads and a hidden size of $768$ was trained for $150$ epochs using the Adam optimizer with a learning rate of $10^{-4}$, $\beta_1 = 0.9$ and $\beta_2 = 0.999$.
Table~\ref{tab:training_acc} shows the accuracy results for \ACC and several other data formats.
The accuracy results for the data formats other than \ACC in Table~\ref{tab:training_acc} were obtained from the work by Zhang et al.~\cite{zhang2022fast}.
For \ACC, we used the exact same training parameters for fair comparison. 
It can be seen that \ACC can provide comparable validation accuracy to FP32 training for all benchmarks. 
All other reported data formats except for INT8 ($2{-}5\%$ accuracy degradation) achieve similar accuracy. 

\subsection{Performance, Power, and Area Evaluation}
\label{sec:ppa}

\renewcommand{\tabcolsep}{3pt}
\begin{table}[t]
    \centering
    \caption{Validation accuracy of Mirage and various data formats~\cite{zhang2022fast}.}
    % \vspace{-0.1in}
    \begin{tabular}{cccccccccc}
    \toprule
    Model &  Mirage & FP32 & bfloat16 & INT8 & INT12& HFP8 & FMAC \\ 
    % \hline
    \cmidrule(lr){1-1}
    \cmidrule(lr){2-2}
    \cmidrule(lr){3-3}
    \cmidrule(lr){4-4}
    \cmidrule(lr){5-5}
    \cmidrule(lr){6-6}
    \cmidrule(lr){7-7}
    \cmidrule(lr){8-8}
    ResNet18 &68.51 & 68.6 & 68.55  & 65.53 & 68.51 &  68.53 & 68.52 \\
    ResNet50 & 75.15 & 75.17 & 75.12  & 71.01 & 75.03 &   75.07 & 75.11 \\
    MobileNet v2 & 68.20 & 68.27 & 68.22  & 65.97 & 68.16 &   68.11 & 68.17 \\
    VGG16 &69.5 & 69.74 & 69.71 & 64.5 & 69.33  & 69.62 & 69.78 \\
    YOLO v2 &73.2& 73.36 & 73.32  & 61.12 & 73.07  & 72.88 & 73.28 \\
    Transformer  &35.4& 35.41 & 35.39  & 29.18 & 35.27 & 35.38 & 35.4 \\
    \bottomrule
    % \vspace{-0.25in}
    
    \end{tabular}
    \label{tab:training_acc}
\end{table}

\renewcommand{\tabcolsep}{2.5pt}
\begin{table}[t]
    \centering
    \caption{Performance, power, and area analysis of MAC units}
    % \vspace{-0.1in}
    \begin{tabular}{ccccccccc}
    \toprule
     &  Mirage  &FP32  & bfloat16 & HFP8  & INT12 & INT8 & FMAC \\ 
     % \cmidrule(lr){1-1}
    \cmidrule(lr){2-2}
    \cmidrule(lr){3-3}
    \cmidrule(lr){4-4}
    \cmidrule(lr){5-5}
    \cmidrule(lr){6-6}
    \cmidrule(lr){7-7}
    \cmidrule(lr){8-8}
    pJ/MAC & 0.21  & 12.42 & 3.20 &1.47 &0.71 & 0.42& 0.11\\
    $\text{mm}^2$/MAC &  0.12 &  9.6E-3 & 3.5E-3 & 1.4E-3 & 7.7E-4 & 4.1E-4 & N/A \\
    $f (Hz)$ &10G& 500M & 500M & 500M  & 1G & 1G & 500M \\
    \bottomrule
    % \vspace{-0.35in}
    
    \end{tabular}
    \label{tab:e_mac}
\end{table}

\begin{figure*}[t]
    \centering
    
     \includegraphics[width=\linewidth]{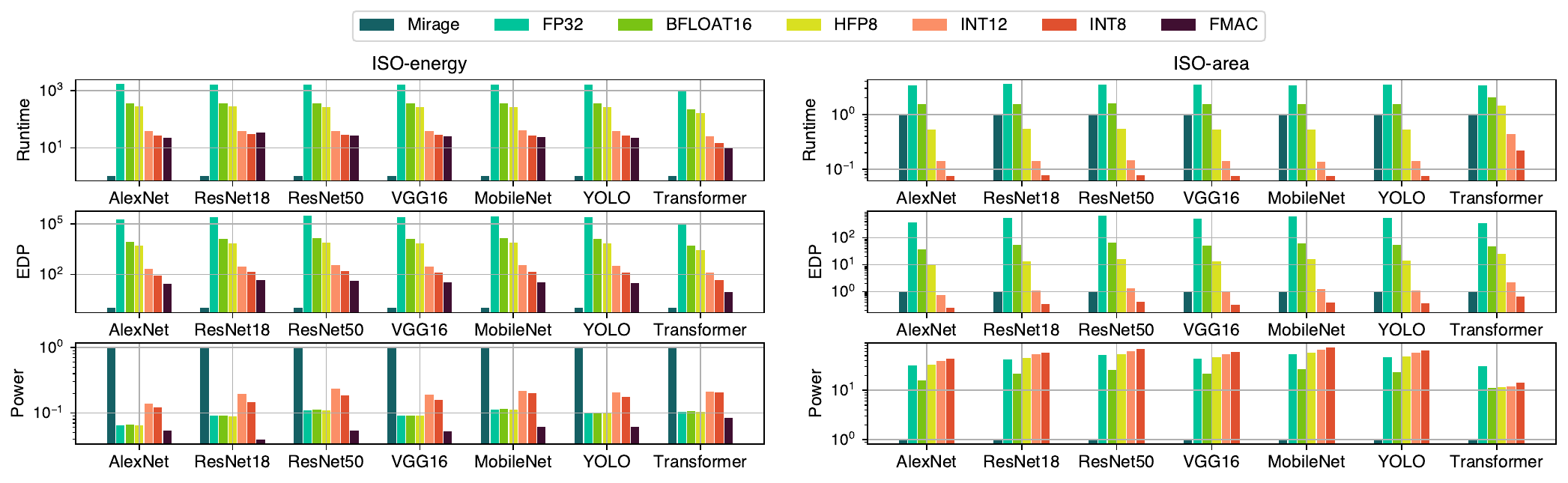}
      % \vspace{-0.1in}
    \caption{Normalized training runtime, EDP and power comparison of \ACC  (eight $16\times 32$ arrays) against systolic arrays using MAC units with various data formats. The plots on the left-hand side show the iso-energy results where the number of MAC units in the systolic arrays is scaled to consume the same energy per MAC operations using the numbers in Table~\ref{tab:e_mac}. The plots on the right-hand side show iso-area results where the number of MAC units in the systolic arrays is scaled to take up the same area as \ACC. As we do not have the area footprint of the FMAC units, we do not show the FMAC numbers in the iso-area results.   
    % \vspace{-0.2in}
    }
    \label{fig:hw_perf}
\end{figure*}

Table~\ref{tab:e_mac} compares the energy consumption and area per MAC operation and clock rate for \ACC's RNS-MMVMUs against systolic arrays with various data formats.
While we could synthesize the digital INT units at $1$ GHz, FP units are typically more complex circuits with longer critical paths than INT MAC units forcing us to reduce the clock frequency to $500$ GHz. 
Compared to other data formats, the main advantage of \ACC is the high clock speed of $10$ GHz with comparable or less energy consumption per MAC. 
In addition to its speed advantage, \ACC also provides a lower energy consumption per MAC ($2{-}59.1\times$) compared to all data formats besides FMAC~\cite{zhang2022fast} (${\sim}2\times$ higher).
% {\color{magenta}

While optical MAC units can reach higher speed and energy efficiency, they typically fall short in computational density as optical devices such as phase shifters and MRRs have a much larger area footprint than digital CMOS gates. 
Therefore, \ACC has a larger area footprint per MAC operation and is less area-efficient than its electronic counterparts.
% }

Fig.~\ref{fig:hw_perf} compares the hardware performance of \ACC against systolic arrays with MAC units using various data formats when training various DNNs.
In Fig.~\ref{fig:hw_perf}, the energy/power consumption of systolic arrays only consists of MAC units while for \ACC, we consider the energy/power consumption of lasers, photonic devices, TIAs, DACs and ADCs, RNS and BFP conversion units, and FP32 accumulators.
The analyses in the figure include iso-energy (per MAC) designs (left) and iso-area designs (right). 
We report results for \ACC with $8$ RNS-MMVMUs, each with $3$ $16\times 32$ MMVMUs (See Section~\ref{sec:sensitivity-analysis} for the justification of the design choices). 
For the iso-energy analysis, we scaled the number of MAC units in systolic arrays to match the energy consumption per MAC operation of \ACC for different data formats using the numbers in Table~\ref{tab:e_mac}.
Similarly, in the iso-area analysis, we increase the number of MAC units in systolic arrays to take up the same area as \ACC. 
We observed that increasing size leads to long latencies to load up the new tile and causes the systolic array performance to go down significantly. 
To avoid this performance drop in systolic arrays, while increasing the number of MAC units, we kept the $16\times 32$ array size fixed and used multiple systolic arrays instead. 
While INT8 cannot meet the high accuracy criteria, it is shown for completeness (See Section~\ref{sec:accuracy-eval}). 

In the iso-energy analysis, the best-performing data format among the systolic array designs is FMAC~\cite{zhang2022fast}. 
Given the same energy per MAC budget, on average across the reported DNNs, \ACC achieves a $23.8\times$ lesser runtime and $32.1\times$ lower EDP than the systolic array with FMAC units. 
However, in this case, \ACC consumes $17.2\times$ higher power consumption.
Compared to the systolic array with FP32 MAC units, on average, \ACC provides $3.5\times$ lesser runtime and $521.7\times$ lower EDP while consuming $42.8\times$ less power.

The iso-area results show that the most efficient datatype that achieves high accuracy, INT12, achieves $5.4\times$ better runtime than \ACC on average due to the large area footprint of \ACC. 
However, while being slower in the iso-area scenario, \ACC has $42.8\times$ lower power consumption and $1.27\times$ lower EDP compared to INT12.
\ACC has $3.5\times$ lesser runtime, $521.7\times$ lower EDP and $42.8\times$ lower power consumption compared to FP32 for the iso-area scenario.

Overall, the results indicate that there exists a tradeoff between runtime, area, and power consumption. 
Compared to digital systolic arrays, given the same energy budget, \ACC can perform faster DNN training, however comes with a higher power consumption and area footprint. 
In contrast, given the same area budget, \ACC has lower power consumption with comparable or better EDP. 

\begin{figure}[t]
    \centering
    % \vspace{-0.1in}
    \includegraphics[width=\linewidth]{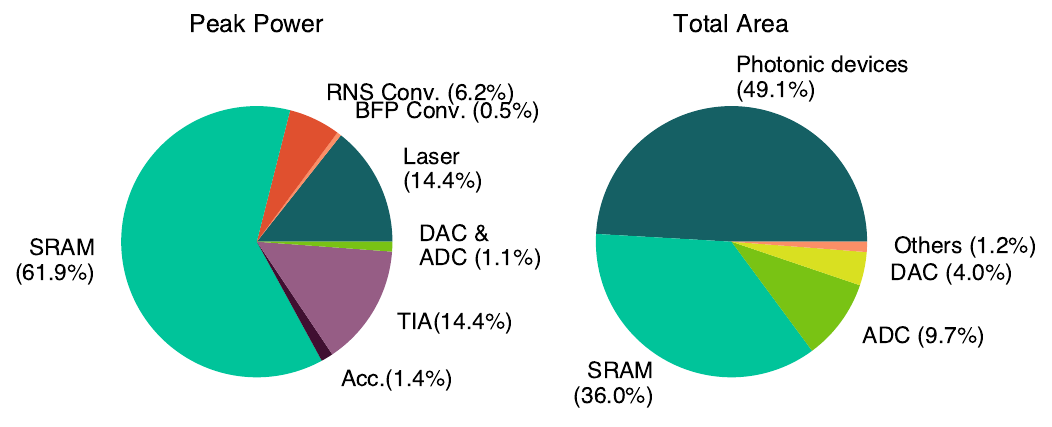}\vspace{-0.2in}
    \caption{Peak power consumption and area breakdown for Mirage. The total peak power consumption is $19.95$ W and the total area is $476.6 \text{mm}^2$.
    \vspace{-0.3in}
    }
    
    \label{fig:breakdown}
\end{figure}

Fig.~\ref{fig:breakdown} shows the peak power and area breakdown for \ACC.
It can be seen that SRAM accesses consume most of the power ($61.2\%$) in \ACC. 
This is mainly because we store all data in FP32 and perform frequent SRAM operations.
To reduce this cost, more efficient data formats (FP16, BFP, etc.) can be chosen to store data and perform nonlinearities---which would reduce the total data storage requirements and the energy consumption per SRAM access.
%We leave the impact of low-precision nonlinear operations on accuracy and hardware performance as part of our future work.  
% {\color{magenta}
It is also noteworthy that in our design, data converters consume only $1.1\%$ of the overall power consumption--which is contrary to a typical analog accelerator where data converter power consumption is a dominating component. 
This is mainly because the reduced bit-precision of DACs/ADCs results in an exponential decrease in their power consumption. 
While the decreasing bit-precision also reduces the required SNR during analog operations, the increased phase shifter length and optical loss prevent the laser power from going down exponentially.
In addition, the power consumption of other components (SRAM arrays, TIAs, accumulators, etc.) increases due to the increasing component count with the use of multiple moduli.
This results in a significant reduction in the relative contribution of data converter power. 
% }

Fig.~\ref{fig:breakdown} (right) shows that most of the area is occupied by photonic devices and SRAM. 
All the components take up $476.6\ \text{mm}^2$ in total, $234\ \text{mm}^2$ for the photonic and $242.7\ \text{mm}^2$ for the electronic chiplet. 
As the photonic and electronic chiplets are stacked via 3D integration, the total area can be considered as the largest of the two chiplets ($242.7\ \text{mm}^2$).

\subsection{\ACC As An Inference Accelerator}
\label{sec:mirage-inference}
The focus of this paper is DNN training.
However, \ACC can also be used to accelerate DNN inference as inference operations are a subset of training operations. 
For completeness, we compare \ACC's performance while running DNN inference against existing photonic and electronic DNN inference accelerators. 
This comparison is shown in Table~\ref{tab:inference}.
\ACC achieves a better throughput in terms of inferences per second (IPS) than all accelerators (by $1.12{-}8.4\times$ compared to photonic and by $176{-}1,856\times$ compared to electronic accelerators) except for ADEPT ($3.37\times$ slower) and TPU v3 ($3.12\times$ slower). 
\ACC provides a better power efficiency (IPS per Watt) than all photonic (by $2.1{-}15.4\times$) and electronic (by $1.74{-}84.7\times$) accelerators except for ADEPT ($2.48\times$ lower). 
% {\color{magenta}
\ACC is more area-efficient (IPS per mm$^2$) than all photonic (by $1.03{-}4.36\times$) and electronic (by $2.38{-}93.9\times$) accelerators except for ADEPT and HolyLight ($1.16\times$ and $8.32\times$ lower, respectively).
% }
It should be noted that these photonic inference accelerators in Table~\ref{tab:inference} provide a much lower dynamic range than \ACC (${\sim}8$ bit vs. ${\sim}15$ bit in \ACC). 
Even for DNN inference, these works can typically can achieve high accuracy only through quantization-aware training~\cite{demirkiran2023electro}. 
With the same methods applied, in \ACC, a lower $b_m$ and a moduli set with a much smaller $M$ can be utilized, resulting in significantly better hardware performance.
% With a hardware configuration better suited to DNN inference, the hardware performance of \ACC can be significantly improved.
%This, however, is out of our scope in this paper as we focus on DNN training. 

\renewcommand{\tabcolsep}{2pt}
\begin{table}[t]
\caption{Mirage vs DNN Inference Accelerators.
%Mirage is customized for training. Inference analysis is provided for completeness
}
\label{tab:inference}
\centering
% \vspace{-0.05in}
\begin{tabular}{ccccccc}
\toprule
 &  \multicolumn{3}{c}{ResNet50} & \multicolumn{3}{c}{AlexNet} \\ 

    \cmidrule(lr){2-4}
    \cmidrule(lr){5-7}
   
 Accelerator & \multicolumn{1}{c}{\begin{tabular}[c]{@{}c@{}}IPS\end{tabular}} & \begin{tabular}[c]{@{}c@{}}IPS/W\end{tabular}& \begin{tabular}[c]{@{}c@{}}IPS/mm$^2$\end{tabular} & \multicolumn{1}{c}{\begin{tabular}[c]{@{}c@{}}IPS\end{tabular}} & \begin{tabular}[c]{@{}c@{}}IPS/W\end{tabular}& \begin{tabular}[c]{@{}c@{}}{IPS/mm$^2$}\end{tabular} \\ 
 % \cmidrule(lr){2-2}
 %    \cmidrule(lr){3-3}
 % \cmidrule(lr){4-4}
 %    \cmidrule(lr){5-5}
 %    \cmidrule(lr){6-6}
   \midrule
\begin{tabular}[c]{@{}c@{}}\textbf{Mirage}\end{tabular} & \multicolumn{1}{c}{10,474} & 1,540.6 & 43.2&\multicolumn{1}{c}{64,963} & 1,904.5 &  267.67\\ 
ADEPT & \multicolumn{1}{c}{35,698} & 1,587.99 &  50.57& \multicolumn{1}{c}{217, 201} & 7,476.78 &307.64\\ 
Albireo-C~\cite{shiflett2021albireo} & \multicolumn{1}{c}{N/A} & N/A & N/A & \multicolumn{1}{c}{7,692} & 344.17 & 61.46\\ 
DNNARA~\cite{peng2020dnnara} & \multicolumn{1}{c}{9,345} & 100 & 42.05 &  \multicolumn{1}{c}{N/A} &  N/A & N/A  \\ 
HolyLight~\cite{liu2019holylight} & \multicolumn{1}{c}{N/A} & N/A & N/A &\multicolumn{1}{c}{50,000} &  900 & 2,226.11\\ 
 \midrule
 Eyeriss~\cite{Eyeriss2017} &  \multicolumn{1}{c}{N/A} & N/A & N/A &\multicolumn{1}{c}{35} & 124.80 &  2.85\\ 
Eyeriss v2~\cite{eyeriss_v2} &  \multicolumn{1}{c}{N/A} & N/A & N/A &\multicolumn{1}{c}{102} & 174.80 & N/A \\ 
TPU v3~\cite{tpuv3} &  \multicolumn{1}{c}{32,716} & 18.18 &  18.00 &\multicolumn{1}{c}{N/A} & N/A & N/A \\ 
UNPU~\cite{unpu} &  \multicolumn{1}{c}{N/A} & N/A & N/A &\multicolumn{1}{c}{346} & 1,097.50 & 21.62\\ 
Res-DNN~\cite{samimi2020} &  \multicolumn{1}{c}{N/A} & N/A & N/A &\multicolumn{1}{c}{386.11} & 427.78 & N/A\\ \hline

\end{tabular}
% \vspace{-0.3in}
\end{table}

\subsection{Managing Noise and Process Variations in \ACC}
\label{sec:noise-mng}

In photonic cores, analog noise, process variations, and fabrication errors can cause noise and errors in the residues.
While our accuracy experiments only consider errors due to tiling and quantization, we take shot noise, thermal noise, and all the optical losses along the path into consideration during our PPA analysis and back calculate the required laser power to achieve the desired bit precision (See Section~\ref{sec:evaluation_method} for details).

% {\color{red}
In addition to these noise sources, process variations can cause bias in phase shifters and drifts in the resonant wavelength in MRRs resulting in errors during operations. 
Prior works proposed various methods such as careful parameter choices during fabrication~\cite{sunny2021crosslight, 9634115mrr}, novel device modifications~\cite{Song:22}, and error correction methods~\cite{Bandyopadhyay:21, PhysRevApplied.18.024018} for MRR- and MZI-based designs to minimize or calibrate away these errors.
These methods can be leveraged in \ACC similar to other photonic hardware.
Moreover, previous works show that increasing DAC precision helps encode values more precisely and reduces the encoding errors in the photonic devices to achieve the desired bit precision~\cite{demirkiran2023electro}.   
The output precision ($b_\text{out}$) in an MDPU is limited by how well one can encode input values onto phase shifters and MRRs.
The total error at the output of an MDPU can be calculated by considering all errors accumulated along the optical path as the signal passes through photonic devices. 
There are $h$ MMUs in an MDPU. Each MMU includes a group of phase shifters representing a $\lceil\text{log}_2m \rceil$-bit value and $2\lceil\text{log}_2m \rceil$ MRRs controlling the route of the signal. 
The precision at the output of an $h$-long MDPU can then be quantified by adding the errors in quadrature as
\begin{equation}
    \Delta\Phi_\text{out} = \sqrt{h\Delta\varepsilon_\text{PS}^2+2h \lceil\text{log}_2m \rceil \Delta\varepsilon_\text{MRR}^2}\ ,
\end{equation}
where $\Delta\varepsilon_\text{PS}$ is the encoding error per phase shifter (phase shifters within a single MMU are considered together) and $\Delta\varepsilon_\text{MRR}$ is the encoding error per MRR.  $\Delta\Phi_\text{out}$ is calculated for the worst-case scenario where the light goes through all the phase shifters.
% \leq 2^{-b_\text{DAC}}$
% \leq 0.3\%$
% with $h$ elements is limited by the phase encoding error $\epsilon_{\Phi}\leq2^{-b_\text{DAC}}$ per element and $\epsilon_\text{MRR}\leq0.3\%$ error per MRR (25 dB extinction ratio)~\cite{Ohno:21}, i.e., $2^{-b_\text{out}}\leq \sqrt{h \epsilon_{\Phi}^2 + 2h \lceil\text{log}_2m \rceil \epsilon_\text{MRR}^2}$.
It should be noted that $\Delta\varepsilon_\text{PS}$ 
and $\Delta\varepsilon_\text{MRR}$ quantities should be measured in the fabricated silicon photonic wafer to precisely estimate the compound noise and the error margins during analog operations. 
However, for this study, we can conservatively assume $\Delta\varepsilon_\text{PS}\leq 2^{-b_\text{DAC}}$ and $\Delta\varepsilon_\text{MRR}\leq 0.3\%$~\cite{Ohno:21}. 
To achieve a $b_\text{out}$-bit output precision, $ \Delta\Phi_\text{out}\leq 2^{-b_\text{out}}$ should be guaranteed. 
Our calculations show that $b_\text{DAC}\geq8$ satisfies this inequality for $b_\text{out}\geq \text{log}_2m$ when $h=16$, which is adequate for achieving high accuracy in \ACC. 
%using 8-bit DACs (instead of 6-bit) satisfies this inequality and can tolerate the encoding errors in \ACC.
Given that DACs and ADCs together consume only $1.1\%$ of the overall power and DACs are used only when a new tile is loaded to the phase shifters, slightly increasing DAC precision (from 6-bit to 8-bit) does not cause a significant change in the overall power consumption. 
With 8-bit DACs~\cite{8bDAC_7062924}, energy and average power consumption of \ACC increases only by $1.09\times$ on average among the evaluated DNNs compared to \ACC with 6-bit DACs. 
% }

Lastly, redundant RNS (RRNS) can be used for error detection and correction in RNS-based systems. 
Demirkiran et al.~\cite{demirkiran2023blueprint} show that by adding redundant moduli to the original set, we can recover from accuracy loss during RNS-based DNN inference in the presence of noise. 
In RRNS, the operations are performed for all moduli as regular RNS.
The errors can then be detected and corrected through majority logic decoding. 
Adding redundant moduli to the set increases the power and area roughly linearly with the number of moduli as the number of components scales linearly with the number of moduli, while throughput stays the same. 
%For example, power consumption and area footprint are expected to increase by $1.33\times$ for one and $1.66\times$ for two redundant moduli.
%We leave the fault tolerance integration as part of our future work. 

% RNS is susceptible to noise as small errors in the residues scale up during output reconstruction, leading to larger errors in BNS.
% When noise is high, this can cause a degradation in accuracy.
% The impact of thermal fluctuations can be minimized through stabilization circuits in MRRs. 
% In \ACC, considering our power analyses, we expect this thermal stabilization cost to be negligible. 
% Moreover, previous works showed that process variations in optical devices can be minimized when designed carefully~\cite{sunny2021crosslight}. 

\section{Related Work}
\label{sec:related-work}

Over the years, many photonic DNN accelerators have been proposed, which almost exclusively target DNN inference. 
Some of these inference accelerators are based on MRR weight banks~\cite{mehrabian2018pcnna}, MZI arrays~\cite{demirkiran2023electro, shen2017deep}, and a mixture of MZIs and MRRs~\cite{shiflett2021albireo, shiflett2020pixel}.
% {\color{cyan}
DNNARA~\cite{peng2020dnnara}, similar to \ACC, uses RNS for performing multiplication and addition operations.
Unlike \ACC, DNNARA does not use any analog property to encode information, instead it uses photonics only to map input operands to the outputs via a one-hot encoded mapping built by using $2{\times} 2$ switches.
This network of switches change the route of the light uniquely to the combination of two operands (i.e., the activated input port and the states of the switches). 
However, each multiply and add operation requires a separate network, resulting in $O(m\text{log}m)$ switches \emph{per operation} for a modulus $m$. 
Although parallelism can be imposed using wavelength division multiplexing (WDM), the number of devices increases rapidly for larger moduli and as more operations are performed in parallel.
In \ACC, each MAC operation requires fewer optical devices ($O(\text{log }m)$), providing a more scalable approach. 
% }

% {\color{orange}
Res-DNN~\cite{samimi2020} and RNSnet~\cite{salamat18} also proposed using RNS to accelerate DNN computation in digital accelerators. 
While these accelerators reported promising results compared to other electronic accelerators, unlike \ACC, they are still bound by the speed of electronic operations which cannot match the high bandwidth of photonics. 
% }
% \revCommon{
In addition, these RNS-based works propose staying in the RNS domain throughout the whole inference. 
While this idea is promising for reducing the RNS-BNS conversion overhead, it has several drawbacks. 
First, it requires performing periodic scaling operations in the RNS domain to stay in the RNS range. 
Second, nonlinear operations cannot be performed using RNS. 
This necessitates using approximations (e.g., Taylor series expansion) for nonlinear operations---which can lead to accuracy loss, especially in training. 
Lastly, staying in the RNS domain for the whole DNN computation limits us to use only integer arithmetic. 
In \ACC, using BFP enables us to preserve the dynamic range during operations and significantly improves the success of DNN training in analog hardware.
Even for inference, these works~\cite{salamat18, samimi2020} use significantly higher input/weight precision than \ACC to achieve high accuracy ($\geq 16$-bit vs. $5$-bit in \ACC). 
Considering that RNS-BNS conversions only consume $6\%$ of the power in \ACC, we believe that the hybrid arithmetic (RNS and FP) approach we propose is a better fit for DNN training. 
% }

% {\color{magenta}
Similar to photonics, ReRAM-based processing-in-memory (PIM) designs also suffer from precision limitations. 
To overcome this issue, some prior works used multiple low-bit cells to compose higher-bit results. 
For example, PRIME~\cite{prime7551380} uses two 3-bit cells to achieve 6-bit precision. 
Another example, PipeLayer\cite{pipelayer7920854}, uses four 4-bit cells to achieve 16-bit precision through shift-and-add operations for DNN inference and training. 
While this approach is similar to using RNS in terms of composing high-precision from low-precision arithmetic, each $b$-bit MAC still produces $\geq$ $2b$-bit result. 
In RNS, bit precision does not grow during operations. 
Compared to PipeLayer, \ACC is $14.4\times$ more power-efficient (OPs/s/W) while being $ 8.8\times$ less area efficient (OPs/s/mm$^2$).  
% } 

Lastly, a few previous works demonstrated fully optical~\cite{bandyopadhyay2023photonic} and hybrid in-situ DNN training~\cite{pai2023experimentally, hughes2018training}. 
Other works combined photonics with alternative training schemes such as direct feedback alignment~\cite{Filipovich:22} and genetic algorithm~\cite{zhang2021efficient}.
While these works are promising in terms of showing the applicability of the photonic technology in DNN training, the demonstrations have been limited to DNNs with only a few layers, small datasets, and simple classification problems so far. 
To train more complex state-of-the-art DNNs, higher precision will be needed---encouraging us to look for innovative accelerator designs. 
In \ACC, by utilizing BFP and RNS, we propose a way of extending the applicability of photonic training accelerators to today's commonly used DNNs.

% There exists digital DNN accelerators~\cite{rns-net, res-dnn} that use RNS for energy-efficient computation and show that it is possible to stay in the RNS domain (without reverting back to binary or decimal) for the entire inference. 
% However, this approach requires overflow detection and scaling and also uses approximations for non-linear operations.
% Most state-of-the-art DNNs today comprise a wide variety of non-linear operations. 
% Using approximations and fixed-point-only arithmetic for these operations can severely degrade the accuracy. 
% Therefore, in our work, we use RNS only for MVM operations and switch back to floating-point arithmetic for non-linear operations.
\section{Discussion}
\label{sec:discussion}

% \revCommon{
% Our studies show that \ACC, an RNS-based photonic accelerator, is inherently less compact than a traditional non-photonic analog design due to the modular operations and more than $2\pi$ phase shifts per MAC. 
% However, the reduced precision during operations still enables us to have a power-efficient design given a reasonable area budget. The high-precision MAC operations that can be achieved only through RNS enables successful training which cannot be obtained via traditional analog design.
% }

% {\color{blue} 
%In our paper, we argued that high-precision ADCs are impractical due to their high energy consumption. 
Over the years, researchers have developed many ADC designs including traditional $\Delta \Sigma$ ADCs, Flash ADCs, Successive Approximation Register (SAR) ADCs, and hybrid ADCs.
In recent years, the architecture trends have shifted mainly towards hybrid and SAR-based architectures combined with techniques such as pipelining and time-interleaving to push the envelope further. 
%The relentless optimization and technology scaling enabled today's data converters to operate close to practical limits. 
Typically, for low-speed ADCs ($f_s{\leq}10^8$ samples/sec), a widely accepted limit is the minimum possible energy spent on a class-B switch capacitor circuit, i.e., $E_\text{ADC}\geq8kT\times$SNR, whereas most high-speed ADCs are technology-limited~\cite{murmann2022introduction}. 
%It should be noted that the energy consumption of data converters is highly architecture and application-dependent. 
This indicates that there can be room for improvement in high-speed data converters with further technology scaling.
However, with technology scaling significantly slowing down, specialization and new designs can only provide limited opportunities to improve energy efficiency. 
Therefore, we believe that our work holds an important role in terms of enabling energy-efficient next-generation analog hardware. 
% }

While \ACC uses photonic technology, the idea of using RNS applies to other analog technologies that suffer from precision restrictions, such as ReRAM, PCM, etc.~\cite{demirkiran2023blueprint}.
For such accelerators, GEMM operations can be performed as is and analog modulo operations can be implemented through electrical analog circuits such as ring oscillators~\cite{ordentlich2018modulo}.
\section{Conclusion}
\label{conclusion}
In this work, we proposed \ACC, an RNS-based photonic accelerator for DNN training.
By combining RNS and BFP, \ACC can successfully train state-of-the-art DNNs at least $23.8\times$ faster and with $32.1\times$ lower EDP in an iso-energy scenario and with $42.8\times$ lower power consumption in an iso-area scenario than systolic arrays. 
Overall, we believe that combining analog computing with RNS is a promising solution to overcome the precision challenges in photonic DNN accelerators.

\section*{ACKNOWLEDGEMENTS}

This work was in part supported by the IARPA MicroE4AI program and by Lightmatter through an internship.

% \section*{Acknowledgements}

%%%%%%% -- PAPER CONTENT ENDS -- %%%%%%%%

%%%%%%%%% -- BIB STYLE AND FILE -- %%%%%%%%
\bibliographystyle{IEEEtranS}
\newpage
\bibliography{refs}
%%%%%%%%%%%%%%%%%%%%%%%%%%%%%%%%%%%%

\end{document}